\documentclass{jfm}
\usepackage{graphicx}
\usepackage{epstopdf, epsfig}
\usepackage{newtxtext}
\usepackage{newtxmath}
\usepackage{natbib}
\usepackage{hyperref}
\hypersetup{
	colorlinks = true,
	urlcolor   = blue,
	citecolor  = black,
}
\usepackage[x11names, svgnames, dvipsnames]{xcolor}
\usepackage{soul} 
\usepackage{float}
\usepackage{lineno}
\usepackage[normalem]{ulem}

\newcommand\revone[1] {{\color{black}{#1}}}
\newcommand\revtwo[1] {{\color{black}{#1}}}

\newcommand\Fro{\mbox{\textit{Fr}}}
\newcommand\Str{\mbox{\textit{St}}}

\shorttitle{Stratified slender body wakes}
\shortauthor{Ortiz, Nidhan and Sarkar}

\title{The high-Reynolds-number stratified wake of a slender body and its comparison with a bluff-body wake}

\author{Jose L. Ortiz-Tarin\aff{1}, Sheel Nidhan\aff{1}, Sutanu Sarkar\aff{1}\corresp{\email{sarkar@ucsd.edu}}}

\affiliation{\aff{1}Department of Mechanical and Aerospace Engineering, University of California San Diego, CA 92093, USA}	

\begin{document}
\maketitle
\begin{abstract}
The high-Reynolds number stratified wake of a slender body is studied using a high-resolution hybrid simulation. The wake generator is a 6:1 prolate spheroid with a tripped boundary layer, the diameter-based body Reynolds number is $\Rey= U_\infty D/\nu = 10^5$, and the body Froude numbers are $\Fro=U_\infty/ND=\{2,10,\infty\}$. The wake defect velocity ($U_d$) decays following  three stages with different wake decay rates  \citep{Spedding1997} as for a bluff body. However,   the  transition  points among stages do not follow the expected $Nt = Nx/U_\infty$ values. Comparison with the wake of a circular disk in similar conditions \citep{Chongsiripinyo2020} quantifies the influence of the wake generator - bluff versus slender - in stratified flow. 
The strongly stratified $\Fro=2$  wake  is in a resonant state. The steady lee waves strongly modulate the mean flow and, relative to the disk, the  6:1 spheroid (a high aspect ratio shape) wake at $\Fro=2$ shows an earlier transition from the non-equilibrium (NEQ) stage to the quasi two-dimensional (Q2D) stage.
The NEQ-Q2D transition is followed by a sharp increase in the turbulent kinetic energy and horizontal wake meanders.
At  $\Fro=10$,  the start of the NEQ stage is delayed for the spheroid. Transfers between kinetic energy and potential energy reservoirs (both mean and turbulence)  are  analyzed and the flows are compared in phase space (local Froude and Reynolds number as coordinates). Overall, the results of this study  point to  the difficulty of finding a universal framework for stratified wake evolution, independent of the features  of the body,  and provide insights into how buoyancy effects depend on the wake generator. 
\end{abstract}

\section{Introduction}


Due to their low drag coefficients, slender bodies are extensively used in aerospace and naval applications. Multiple studies have described the flow around these bodies focusing on the drag force, the boundary layer, and the flow separation \citep{Wang1970, Costis1989, Wang1990, Chesnakas1994, Fu1994, Constantinescu2002, Wikstrom2004}. However, despite their presence in many underwater applications, only a few works have looked into the wake of a slender body \citep{Chevray1968,Jimenez2010,Kumar2018} and, only recently, the far wake of a slender body has been studied \citep{Ortiz2021}.

The near wake of a slender body with a turbulent boundary layer (TBL) is characterized by having a small recirculation region. The recirculation region is surrounded by a ring of small scale turbulence that emerges from the boundary layer and does not show strong vortex shedding \citep{Jimenez2010,Posa2016,Kumar2018,Ortiz2021}. As a result, the wake is thin and develops slowly compared to the wake of bluff bodies. These particular features of the slender body high-$\Rey$ near wake lead to interesting effects further downstream: (i) despite having a smaller drag coefficient than bluff bodies, the defect velocity ($U_d=U_\infty-U$) of the slender body wake can be larger than that of a bluff body for a long downstream distance, (ii) the turbulent kinetic energy of the wake shows an off-center radial peak at the location where the turbulent boundary layer separates -- instead of a Gaussian profile with a central peak, and (iii)  helical instabilities come into play only in the intermediate and far field of the wake. These particularities affect the scaling laws of the wake. In a domain spanning $80D$ the defect velocity, the kinetic energy, and the dissipation do not follow the classic high-$\Rey$ scaling and they decay differently than bluff body wakes \citep{Ortiz2021} exhibiting a non-equilibrium scaling of dissipation \citep{Vassilicos2015,Dairay2015}. 

The few studies that look into slender body wakes assume that the body moves in an unstratified environment, where the density of the surrounding fluid is constant. However, in a realistic underwater marine environment the effect of density stratification due to salinity and temperature can become relevant. Density stratification suppresses vertical motions, triggers the formation and sustenance of coherent structures, and leads to the radiation of internal gravity waves. More importantly, in a stratified environment, the wake of a submersible lives longer than in an unstratified environment, i.e., it takes more time for the flow disturbance to die out \citep{Spedding2014}. The study of stratified wakes has been nearly exclusively focused on the flow past bluff bodies \citep{Lin1979,Hanazaki1988,Lin1992,Chomaz1992,Orr2015,Pal2017} and underwater topography \citep{Drazin1961,Castro1983,Baines1998}. Here, we study the influence of stratification on the high-$\Rey$ wake of a prolate spheroid with a turbulent boundary layer.

The strength of ambient stratification is measured by the body-based Froude number $\Fro=U_\infty/ND$. This is the ratio between the convective frequency of the flow, $U_\infty/D$ -- where $U_\infty$ is the freestream velocity and $D$ the diameter of the body -- and the buoyancy frequency $N$. In the wake of ocean submersibles, $\Fro \sim O(1-10^2)$. However, since the velocity deficit in the wake $U_d(x)$ decays with the streamwise distance and the wake width $L(x)$ increases, the Froude number defined with local variables $\Fro_l=U_d/NL$ decreases as the flow evolves. Thus, even in a weakly stratified environment, eventually all wakes are affected by stratification.         

Since the relative strength of stratification increases {locally} as the flow develops, the evolution of the stratified wake is multistage in nature. Based on the measurements of $U_d$ and $L$ in high-$\Fro$ (i.e. initially weak stratification) bluff body wakes,  \cite{Spedding1997} identified three regimes in  stratified wake evolution \revtwo{based on the power-law decay rates of $U_d$. These regimes are generally identified by empirically fitting decay rates to $U_d$ in different temporal (or spatial) regions and are as follows:}

\revtwo{
\begin{enumerate}
	\item Three-dimensional (3D) regime: Close to the generator,  wake decay is similar to the unstratified wake of the corresponding body shape. This is the so-called 3D regime  and lasts until the buoyancy time defined by $Nt = Nx/U = x/D \cdot 1/\Fro$  approaches $ O(1)$, equivalently until  $x/D \sim  \Fro$.
		
	\item Non-equilibrium (NEQ) regime: As the wake evolves, buoyancy effects become progressively stronger. The decay of $U_d$ slows relative to the  3D regime and, furthermore,   anisotropy between the vertical and horizontal velocity components increases. \cite{Spedding1997} reported this non-equilibrium (NEQ) region to last for  $Nt \approx 2-50$. Later, temporal simulations of \cite{Brucker2010} and \cite{Diamessis2011} found an increase in the span of the NEQ regime at higher Reynolds numbers. $U_d \sim x^{-0.25 \pm 0.04}$ during the NEQ regime according to \cite{Spedding1997}. However, there has been some variability in the observed NEQ decay rate in later studies. \cite{Bonnier2002} reported a NEQ regime with $U_d \sim x^{-0.38}$ for $1.5 < \Fro < 5$. \cite{Brucker2010} and \cite{Diamessis2011} reported $U_d \sim Nt^{-1/4}$ during the NEQ regime in their temporal simulations. \cite{Chongsiripinyo2020} found that $U_d \sim x^{-0.18}$ during the NEQ regime of their $\Fro = 2$ and $10$ disk wakes. For wakes with $\Fro \sim O(1)$, the NEQ decay rate is preceded by a pronounced oscillatory modulation in $U_d$ which is  linked to  lee waves \citep{Pal2017,Chongsiripinyo2020,Ortiz-tarin2019}.
	
	\item Quasi two-dimensional (Q2D) regime: After the NEQ regime, the stratified wake enters into the quasi two-dimensional regime (Q2D regime). The Q2D regime is characterized by transition in the  $U_d$ power law to a significantly increased  decay rate, e.g.   \cite{Spedding1997} reports a transition to  $U_d \sim x^{-3/4}$.  In the Q2D regime, the wake progressively organizes into vortices that meander primarily in the horizontal plane \citep{Gourlay2001,Dommermuth2002,Brucker2010} and take the form of `pancakes'. Although the wake motion in this regime is primarily in the horizontal plane, there is a variability in the vertical direction in the form of layers \citep{Spedding2002}, hence the prefix `quasi'.
\end{enumerate}

 In recent literature,  stratified wakes have been characterized using turbulence features \citep{Zhou2019, Chongsiripinyo2020} instead of the $U_d$-based criteria of \cite{Spedding1997}.  These studies are motivated by an attempt to connect buoyancy-related wake transitions to the broader stratified turbulence field. 
}



Notice that the arrival of the wake into each of the three stages in its evolution  depends on the value of $Nt$, which is equivalent to a downstream distance of $x/\Fro$  from the wake generator. At high Froude number, the downstream distance required to reach the NEQ and Q2D regions can become very large. Consequently, the size of the computational domain required to access these regimes rapidly becomes computationally unfeasible. To circumvent these limitation temporal simulations were used in the study of \cite{Gourlay2001}. Temporal simulations use a reference frame moving with the wake where time correlates with streamwise distance in a fixed reference frame. By assuming that the streamwise development of the flow is slow, periodic boundary conditions can be used and the equations are advanced in time without the need of introducing the wake generator. This reduces the computational cost significantly. Most of the studies that have contributed to our current understanding of stratified wakes use temporal simulations \citep{Gourlay2001, Dommermuth2002, Brucker2010, Diamessis2011,DeStadler2012, Abdilghanie2013, Redford2015, Rowe2020}.

The main drawback of the temporal model is the influence of its initialization. Since the flow at the wake generator is not solved, the starting profiles of the mean and turbulence have to be assumed. These simulations lack some specific features that are generated due to the body, e.g., steady lee waves, near-wake buoyancy effects, and the vortical structures shed from the boundary layer. Even when it is tempting to assume that body specific features are lost far from the body, the universality of the wake decay has remained elusive to experiments \citep{Bevilaqua1978, Wygnanski1986, Redford2012}, even in unstratified wakes. An alternative to temporal simulations are body inclusive simulations that retain the wake generator dependent features at the expense of a higher computational cost and a limited domain size \citep{Orr2015,Chongsiripinyo2017,Pal2017,nidhan_dynamic_2019,more2021orientation}.  

To the best of authors' knowledge, \cite{Ortiz-tarin2019} performed the first study of a stratified flow past a slender body that investigates the near and intermediate wake dynamics. Their analyses reveal that at $\Fro\sim O(1)$ the type of separation and the subsequent wake establishment is strongly dependent on the characteristic frequency of the lee waves and the aspect ratio of the body. When half the wavelength of the steady lee waves ($\lambda=2\pi \Fro$) matches the length of the slender body ($L$), the separation of the boundary layer is inhibited by buoyancy effects. Based on this condition, a critical Froude number can be defined $\Fro_c=L/D\pi$. When $\Fro>\Fro_c$ stratification suppresses the generation of turbulence in the near wake, when $\Fro\approx \Fro_c$ buoyancy strongly limits the flow separation and can lead to a relaminarization of the wake at low Reynolds numbers. Finally, when $\Fro<\Fro_c$, the lee waves enlarge the separation region and there might be an increase in the turbulence intensities in the wake. When $\Fro\approx \Fro_c$ the wake is in a resonant state with both the separation and the wake dimensions being strongly modulated by the steady lee waves \citep{Hunt1980, Chomaz1993, Ortiz-tarin2019}.   


As mentioned before, the use of body inclusive simulations has one major limitation, i.e., the high computational cost. Due to the high resolution required to solve the boundary layer of the wake generator, the downstream domain is limited and thus the possibility of looking into the far wake gets significantly restricted, particularly at high $\Rey$. \cite{VanDine2018} presented a hybrid spatially-evolving model, which builds  on the  hybrid temporally-evolving model of \cite{Pasquetti2011},  and addresses most of the aforementioned problems. The hybrid method  uses inflow conditions generated from a well-resolved body-inclusive simulation to  perform a separate temporal simulation in the case of \cite{Pasquetti2011} or spatially-evolving simulation in the work of  \cite{VanDine2018} without including the body. By doing so, the amount of required points is substantially reduced since the flow near the body does not have to be resolved. This important reduction of the computational cost allows one to extend the domain farther downstream to gain insight in the far wake.

Here, we use a hybrid method that combines a body-inclusive simulation and a spatially evolving body-exclusive simulation to study the stratified high-$\Rey$ far wake of a slender body for the first time. The Reynolds number is set to $\Rey=U_\infty D/\nu=10^5$ and two levels of stratification are used, $\Fro= U_\infty/ND = 2$ and $10$. The simulation at $\Fro=10$ allows us to study the evolution of a weakly stratified wake in a domain that spans $80D$. Additionally, $\Fro=2$ is chosen because it is close to the critical Froude number for a 6:1 prolate spheroid $\Fro_c= (L/D)/\pi = 6/\pi$. At the critical Froude number, the size of the separation region is strongly reduced by the lee waves \citep{Ortiz-tarin2019}. These choices also allow us to compare our results with the findings of \cite{Chongsiripinyo2020} (hereafter referred as CS20) regarding the stratified wake of a disk.

In CS20, the stratified wake of a disk at $\Rey=5 \times 10^4$ is studied at $\Fro={2,10,50,\infty}$. Apart from a detailed analysis of the decay rates of the mean and turbulent quantities, CS20 links the general evolution of stratified homogeneous turbulence \citep{Brethouwer2007,deBKops2019} with the evolution of the wake turbulence. As the disk wake evolves, the influence of buoyancy is `felt' by the turbulent motions at progressively smaller scales. First the mean flow and the large scales and later the r.m.s. velocities are affected by stratification. Simultaneously, the horizontal eddies start gaining energy. Based on the strength of these effects, three distinct stages can be identified: weakly, intermediate and strongly stratified turbulence. In CS20, the transition between these regimes is examined and parameterized using local Froude and Reynolds numbers. \cite{Zhou2019} also examined these transitions and their link with the evolution of stratified homogeneous turbulence using temporal simulations.

The present work is the continuation of \cite{Ortiz2021} -- referred to as ONS21 -- where the unstratified wake of a 6:1 prolate spheroid with a turbulent boundary layer was studied and compared with a large number of simulations and experiments. In our previous study we found that the particularities of the slender body wake, e.g., small recirculation region, low entrainment, large defect velocity, bimodal distribution of the turbulent kinetic energy, among others affect the wake decay significantly. In this study we are analyzing how these features affect the evolution of the stratified wake. We also analyze the simulations of CS20 to closely compare our results with the stratified bluff body wake. 


Some of the questions we want to answer are: do the stratified decay laws and their transition points depend on the shape of wake generator? how does the turbulence evolve in stratified slender body wakes and are there difference with bluff body wakes? how  does the phase-space evolution of the stratified turbulence compares between bluff and slender body wakes? In broader terms, we attempt to find whether a turbulent stratified wake retains some imprint of the wake generator in the mean and turbulence evolution.

A description of the solver and the methodology is given in \S 2. The wakes are visualized in \S 3. The decay of the mean wake properties is analyzed in \S 4. Finally, the evolution of the turbulence and the phase-space analysis of the wake are presented in sections \S 5 and \S 6, respectively. The study is concluded in \S 7.







\section{Methodology}

%
%
%
%
%
%

To study the far wake of a slender body at a high Reynolds number we use a hybrid simulation. The hybrid model combines two simulations:  body-inclusive (BI) that solves the flow past the wake generator and body-exclusive (BE) that resolves the intermediate and far wakes. 
Here, we use a spatially-evolving simulation following the procedure validated by \cite{VanDine2018}. In the implementation, data from a selected cross-plane in the BI simulation is interpolated on to a new grid and used as an inlet boundary condition for the BE stage. This procedure allows us to alleviate the natural stiffness of the wake problem. Whereas the BI simulation is designed to capture the turbulent boundary layer and the flow separation, the BE simulation resolves the turbulence in the wake. Both the grid size and the time step required to solve the turbulent boundary layer are much smaller than those needed in the intermediate and far wakes. This method leads to significant savings in computational cost without compromising accuracy. 

The setup and the solver here are the ones used in ONS2021 with the addition of stratification. Both simulations solve the three-dimensional Navier-Stokes with the Boussinesq approximation in cylindrical coordinates. The solver uses a third-order Runge-Kutta method combined with second-order Crank-Nicolson to advance the equations in time. Second-order-accurate central differences are used for the spatial derivatives in a staggered grid. A wall-adapting local eddy viscosity (WALE) is used to properly capture the turbulent boundary layer dynamics \citep{Nicoud1999}. Both the BI and BE simulations use Dirichlet boundary conditions at the inflow, convective outflow and Neumann at the radial boundary. Similarly to \cite{Ortiz-tarin2019} a sponge layer is added to the boundaries to avoid the spurious reflection of gravity waves.   

An immersed boundary method \citep{Balaras2004, Yang2006} is used to resolve the flow past a 6:1 prolate spheroid at zero angle of attack. A numerical bump is introduced on the surface of the body to accelerate the transition of the boundary layer to turbulence. The annular bump is located where the surface favorable pressure gradient is nearly zero. This location is found at approximately $0.5 D$ from the nose. The radial extent of the bump is $0.002 D$ ($\sim$ 15 wall units) and the streamwise extent is $0.1 D$. 

The stratification is set by a linear background density profile characterized by the Froude number, $\Fro = U_\infty/ND$, where $N$ is the buoyancy frequency. Three levels of stratification are simulated: $\Fro=2, 10$ and $\infty$. $\Fro=2$ is close to the critical Froude number $\Fro_c = 6/\pi$ for the 6:1 spheroid at which the suppression of turbulence in the wake by stratification is optimal \citep{Ortiz-tarin2019}. $\Fro=10$ is a moderate level of stratification closer to oceanic values. Finally $\Fro=\infty$ is the unstratified case which will be used as a reference (ONS21).   

The cylindrical coordinate system is $(x,r,\theta)$ with the origin at the body center. For convenience, the Cartesian coordinate system $(x,y,z)$ will also be used, where $z$ is the vertical direction aligned with gravity, $y$ is the spanwise direction, and $x$ is the streamwise direction.  

The BI grid is designed to resolve the turbulent boundary layer and the small-scale wake turbulence. The turbulent boundary layer is resolved with $\Delta x^+=40$, $\Delta r^+=1$, and $r\Delta \theta^+=32$. There are 10 points in the viscous sublayer and 130 across the buffer and log layers. \revone{The mean velocities and turbulence intensities within the boundary layer were validated against existing studies \citep{Posa2016,Kumar2018} and the law of the wall.  Additionally a grid refinement study was performed to guarantee the independence of the statistics to the grid choice.}

In the wake, the peak ratio between the grid size and the Kolmogorov length $\eta = (\nu^3/\epsilon)^{1/4}$, in both BI and BE domains is $\max(\Delta x/\eta)=7.5$, $\max(\Delta r/\eta)=6$, and $\max(r \Delta \theta/\eta)=5$. \revone{A figure showing the ratio between the Kolmogorov scale and the grid resolution can be found in ONS21 (figure 2).  Additionally, the unstratified wake decay coincides with all the previous existing numerical and experimental works on slender body wakes (see figure 1 of ONS21).}

The domain size in the stratified cases is large so that internal gravity waves are weak before reaching the sponge region near the walls. The total number of grid points across BI and BE domains is approximately 1.5 billion in the unstratified case and 2 billion in the stratified simulations. Tables \ref{tab:bi} and \ref{tab:be} include the most relevant parameters of BI and BE simulations, respectively. Further details on the grid design can be found in section 2 of ONS21.

\begin{table}
\begin{center}
\begin{tabular}{lccccccccc}
    Case  & $\Rey$   &   $\Fro$ &  $L_r$ & $L_{\theta}$ & $L^-_x$ & $L^+_x$ & $N_r$ & $N_{\theta}$ & $N_{x}$ \\[3pt]
       ~~1   & ~~$10^5$ & ~$\infty$~ & 5 & $2\pi$ & 8 & 15 & 746 &512 & 2560\\
       ~~2   & ~~$10^5$ & ~10 & 60 & $2\pi$ & 20 & 30 & 848 & 512 & 3072\\
       ~~3  & ~~$10^5$ & ~$2$ & 60 & $2\pi$ & 20 & 30 & 848 &512 & 3072\\

  \end{tabular}
 \caption{Parameters of the body-inclusive simulation of prolate 6:1 spheroid. $L_x^-$ and $L_x^+$ are the upstream and downstream distances from the wake generator.} 
{\label{tab:bi}}
\end{center}
\end{table}

\begin{table}
\begin{center}
\begin{tabular}{lccccccccc}
    Case  & $\Rey$   &  $\Fro$ &  $L_r$ & $L_{\theta}$ & $x_{e}$ & $L_x$ & $N_r$ & $N_{\theta}$ & $N_{x}$ \\[3pt]
       ~~1   & ~~$10^5$ & ~$\infty$~ & 10 & $2\pi$ & 6 & 80 & 479 &256 & 4608\\
       ~~2   & ~~$10^5$ & ~10 & 57 & $2\pi$ & 9 & 90  & 619 & 256 & 4608\\
       ~~3  & ~~$10^5$ & ~$2$ & 57 & $2\pi$ & 9 & 90 & 619 & 256 & 4608\\

  \end{tabular}
 \caption{Parameters of the body-exclusive simulations. $x_{e}$ is the extraction location of the BI simulations that is fed as inlet to the BE simulations.}{\label{tab:be}}
 \end{center}
\end{table}

\begin{table}
\begin{center}
\begin{tabular}{lccccccccc}
    Case  & $\Rey$   &  $\Fro$ &  $L_r$ & $L_{\theta}$ & $L_x^{-}$ & $L_x^{+}$ & $N_r$ & $N_{\theta}$ & $N_{x}$ \\[3pt]
       ~~1   & ~~$5\times 10^4$ & ~$\infty$~ & 15.14 & $2\pi$ & 30.19 & 125.51 & 364 &256 & 4608\\
       ~~2   & ~~$5\times 10^4$ & ~10 & 80 & $2\pi$ & 30.19 & 125.51  & 529 & 256 & 4608\\
       ~~3  & ~~$5\times 10^4$ & ~$2$ & 80 & $2\pi$ & 30.19 & 125.51 & 529 & 256 & 4608\\

  \end{tabular}
 \caption{Parameters of the disk simulations (CS20).}{\label{tab:disk}}
 \end{center}
\end{table}

Once the flow has reached statistically steady state, the statistics are obtained by temporal averaging, denoted by $\langle \cdot \rangle$. Instantaneous quantities are written with lower case, mean quantities with upper case, and fluctuations with prime. In the stratified cases the average is performed over $270D/U_\infty$, approximately three flow-throughs. In the unstratified simulation flow statistics are obtained through the temporal (over $100D/U_\infty$) as well as azimuthal averaging. Apart from temporal averaging, some statistics are obtained from cross-wake area integration denoted by $\{ \cdot \}$. Unless otherwise indicated, the integral is performed over a cross-section of radius $4D$. All the flow statistics presented here die out well before they reach the limit of the integrated region. 

Reported velocities and lengths are normalized with the free-stream velocity $U_\infty$ and the body minor axis $D$, respectively. The normalized streamwise distance from the center of the body $x$ is also measured as a function of the buoyancy frequency and the time. The time in the $Nt$ axis refers to time measured by an observer attached to the mean flow that sees the body move at a speed $-U_\infty$. A Galilean transformation yields $x/Fr=Nt$. 
  

To compare the stratified wake of the 6:1 spheroid with that of a bluff body we use the body-inclusive disk wake simulations of CS20. The solver used in CS20 is the same as the one used here although, instead of using the WALE closure model, CS20 uses a variant of dynamic Smagorinsky. The eddy-viscosity model was changed in the spheroid simulations since WALE was demonstrated to capture the behavior of the turbulent boundary layer with the resolution used in the present wall-resolved LES. Both sets of simulations are very well resolved and have a small subgrid contribution -- see ONS21 and CS20 -- hence the validity of the comparison. Further details of the simulations can be found in ONS21 and CS20. The main parameters of disk simulations are shown in table \ref{tab:disk}.

\section{Visualizations} \label{sec:vizDiskSpd}

\begin{figure}
	\centerline{\includegraphics[scale=0.5]{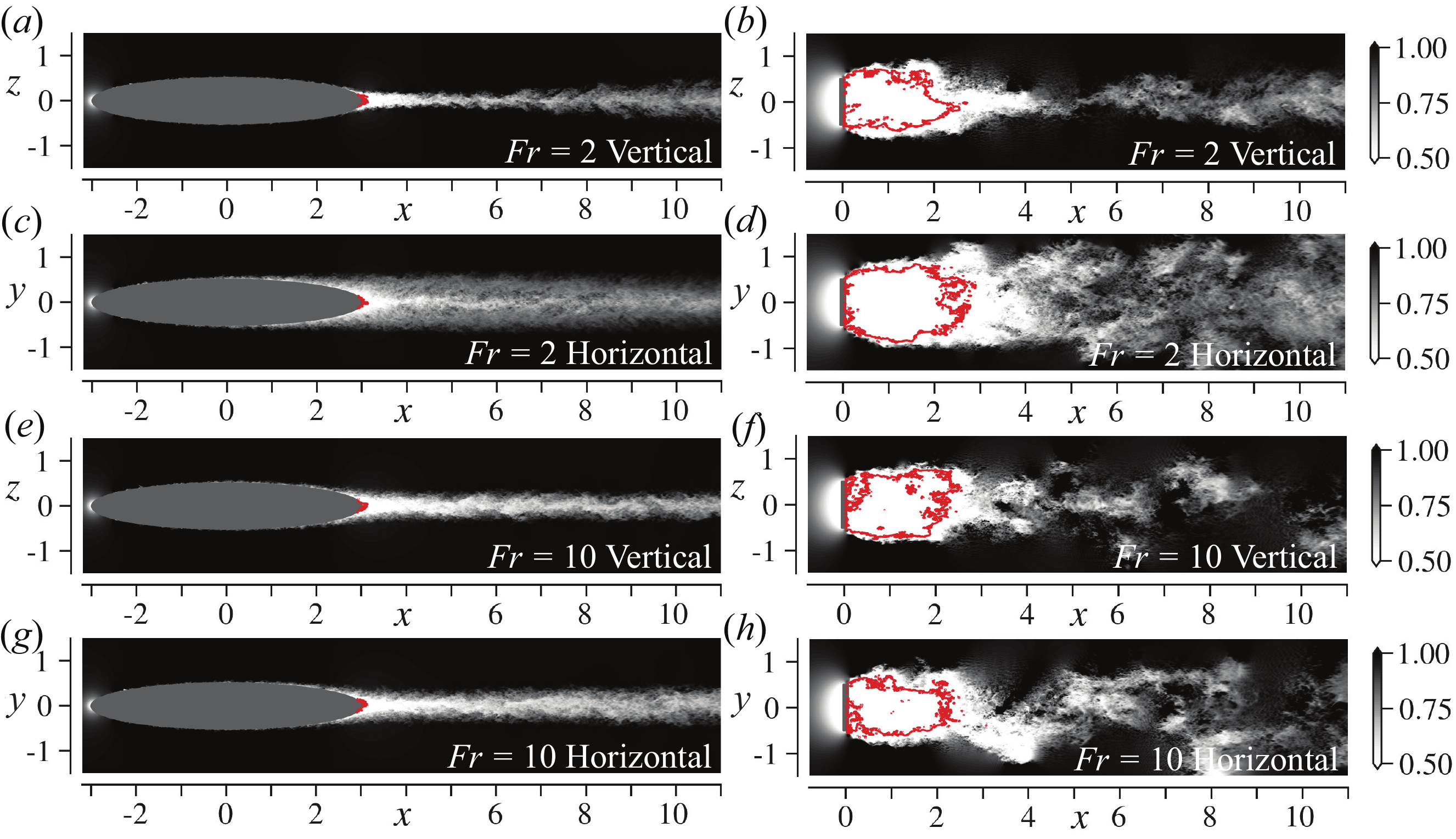}}
	\caption{Instantaneous contours of streamwise velocity in the near wake for spheroid (left) and disk wakes (right) at $\Fro = 2$ and $\Fro = 10$ on center-vertical ($y=0$) and center-horizontal ($z=0$) planes. Red isolines show the limit of recirculation regions where the streamwise velocity is zero.}
	\label{fig:disk_spd_comp}
\end{figure}

Figure \ref{fig:disk_spd_comp} shows instantaneous snapshots of the near wake of a spheroid and a disk at $\Fro = 2$ and $\Fro = 10$. At both $\Fro$, the near wake structure of the two bodies is very different. Compared to the spheroid with turbulent boundary layer (TBL), the disk wake has a large recirculation region ($\sim 2D$), as shown by the red isolines in figure \ref{fig:disk_spd_comp}. This large recirculation region oscillates \citep{Rigas2014} and generates a vortex shedding structure that is advected downstream \citep{Nidhan2020}. In a spheroid with TBL, the recirculation region is very small ($\sim 0.1D$) and is surrounded by the small scale turbulence of the boundary layer. As a result, the near wake is highly organized and large scale oscillations are not observed in the near wake \citep{Jimenez2010,Kumar2018,Ortiz2021}. Only further downstream, does the wake begin to show a helical structure. This change in the structure of the slender body wake has been found to lead to a change in the decay rate and dissipation scaling in the unstratified wake (ONS21). In the following sections, we will analyze how the differences between  the near wake of a disk and that of a spheroid lead to distinct trends of mean and turbulence evolution in a stratified environment. But first, let us describe different snapshots of the spheroid intermediate and far wakes. Snapshots of the disk intermediate and far wakes can be found in CS20.

\begin{figure}
	\centerline{\includegraphics[scale=0.5]{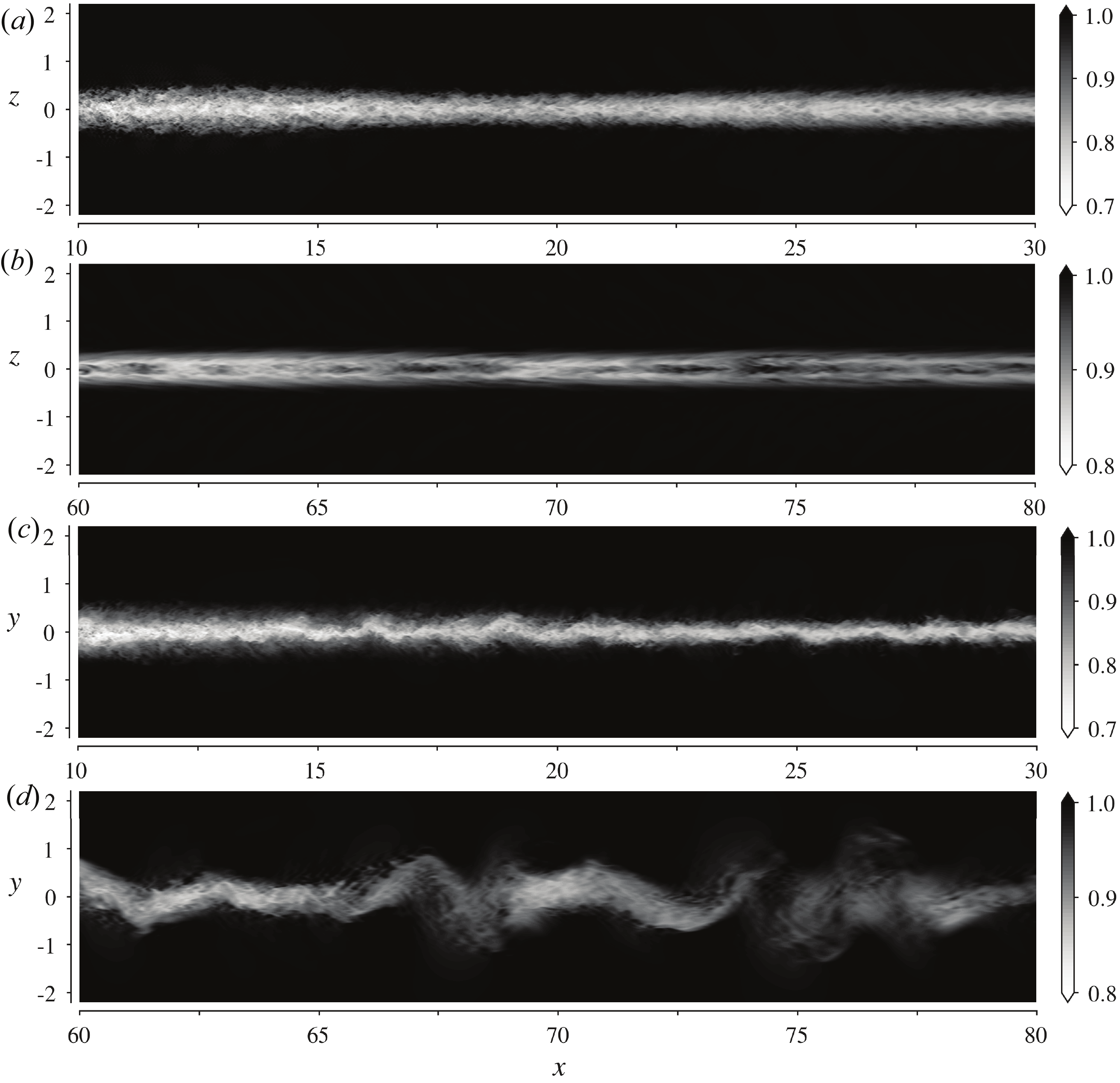}}
	\caption{Instantaneous contours of streamwise velocity of the spheroid $\Fro=2$ wake in center-vertical (a,b) and center-horizontal planes (c,d).}
	\label{fig:fr2_wake_viz}
\end{figure}

Figure \ref{fig:fr2_wake_viz} shows an instantaneous visualization of the spheroid $\Fro=2$ wake in the center-vertical and center-horizontal planes. One of the distinctive features of the spheroid wake is that at $\Fro\sim O(1)$ the separation of the boundary layer can be strongly modulated by the steady lee waves \citep{Ortiz-tarin2019}. This interaction between the lee waves and the wake is particularly strong when the Froude number is close to a critical Froude number $\Fro_c=AR/\pi$, where $AR$ is the body aspect ratio. When $\Fro\approx \Fro_c$, half the wavelength of the lee wave ($\lambda/D=2\pi \Fro$) coincides with the length of the body and the size of the separation region is reduced. The flow is then in what is called resonant or saturated lee wave regime, \citep{Hanazaki1988,Chomaz1992}. At low Reynolds numbers, this effect can lead to the relaminarization of the turbulent wake \citep{Pal_JFMrapids2016, Ortiz-tarin2019}. In the present case, figure \ref{fig:fr2_wake_viz}(a) reveals that, even at $\Rey=10^5$, the wake height is strongly modulated by the waves, although the wake is not relaminarized due to the high $\Rey$ of the flow. For example, the wake height exhibits oscillations with a wavelength of $\lambda/D= 2\pi \Fro = 4\pi$. The modulation of the wake by the waves leads to an unusual configuration in the intermediate wake ($x=20-40$) where the wake width $L_H$ is smaller than the wake height $L_V$, figure \ref{fig:fr2_wake_viz}(a) and (c). In these figures, sinuous oscillations are observed only in the horizontal plane (figure \ref{fig:fr2_wake_viz}d) due to strong stratification. These horizontal sinuous instabilities contrast with the lee wave induced varicose modulation in the vertical plane. As the wake evolves, the $L_H < L_V$ configuration transitions to the expected $L_H > L_V$. In this late region, the small-scale turbulence of the boundary layer has been dissipated and a layered-layer structure is observed in the vertical plane, figure \ref{fig:fr2_wake_viz}(b). The qualitative trends of $L_H$ and $L_V$ discussed here are quantified in \S\ref{sec:meanFlow}.

\begin{figure}
	\centerline{\includegraphics[scale=0.5]{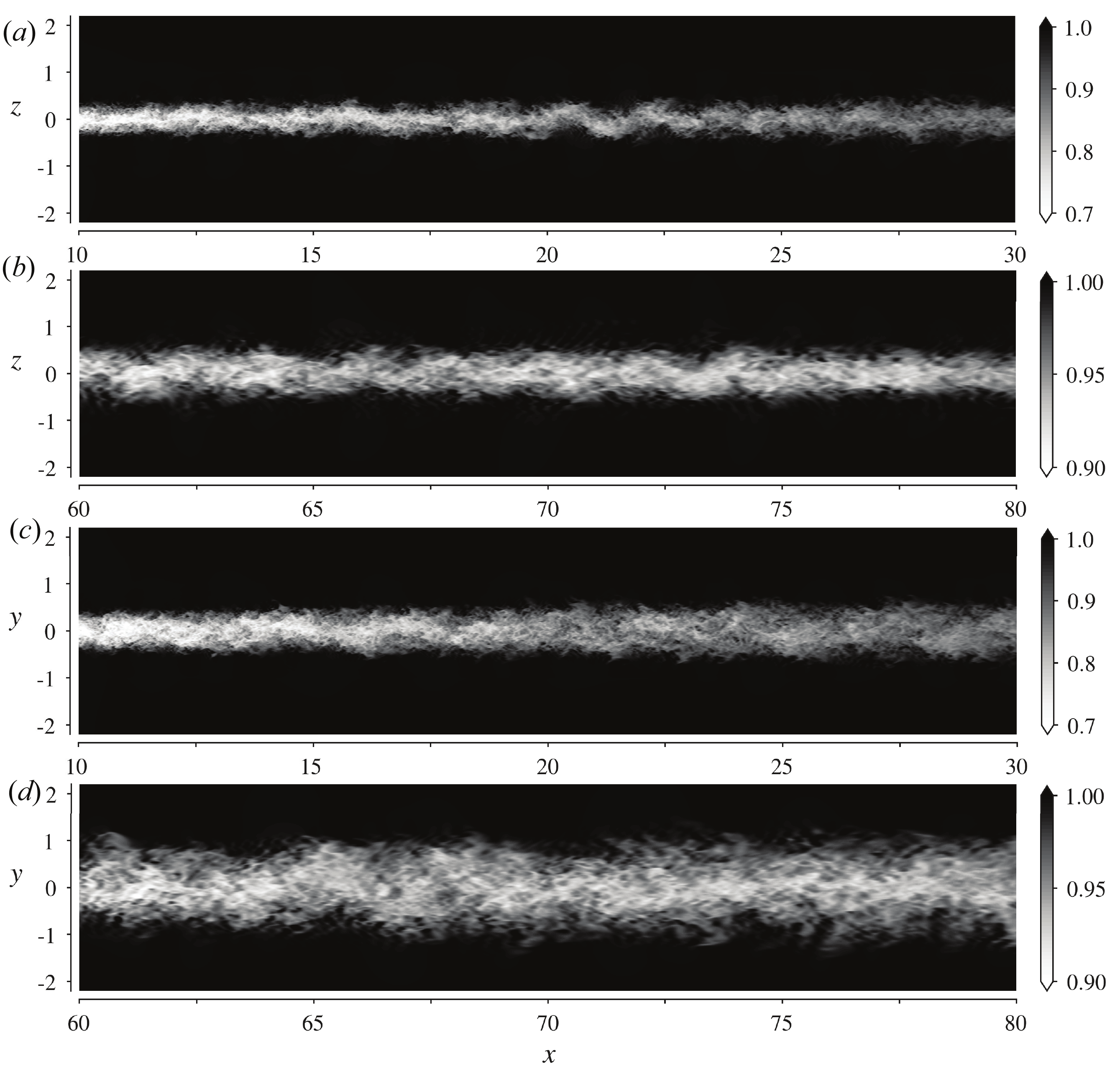}}
	\caption{Instantaneous contours of streamwise velocity of the spheroid $\Fro=10$ wake in center-vertical (a,b) and center-horizontal planes (c,d).}
	\label{fig:fr10_wake_viz}
\end{figure}

The main features of the $\Fro=10$ wake can be observed in the instantaneous snapshots of figure \ref{fig:fr10_wake_viz}. The near wake, figure \ref{fig:fr10_wake_viz}(a,c), is thin and carries the small scale turbulence generated in the boundary layer. Similar to the unstratified wake of ONS21, in the $x<20$ region, it has a quasi-cylindrical structure.  Only after $x\approx 20$, a helical structure develops. In the unstratified wake, the oscillation found at $x\approx 20$ is present until the end of the domain. Here, the $\Fro=10$ wake does not show major oscillations after $x\approx 30$. Stratification restrains the vertical motions in the wake and enhances the horizontal spread as can be seen in the visualization of the late wake, figure \ref{fig:fr10_wake_viz}(b,d). Unlike the $\Fro = 2$ wake, the horizontal and vertical $\Fro = 10$ wake extent grow monotonically  with increasing downstream distance.

\section{Evolution of the mean flow in spheroid and disk wakes} \label{sec:meanFlow}

\subsection{Evolution of the mean defect velocity ($U_d$)} \label{sec:meanUd}

\begin{figure}
	\centerline{\includegraphics[scale=0.5]{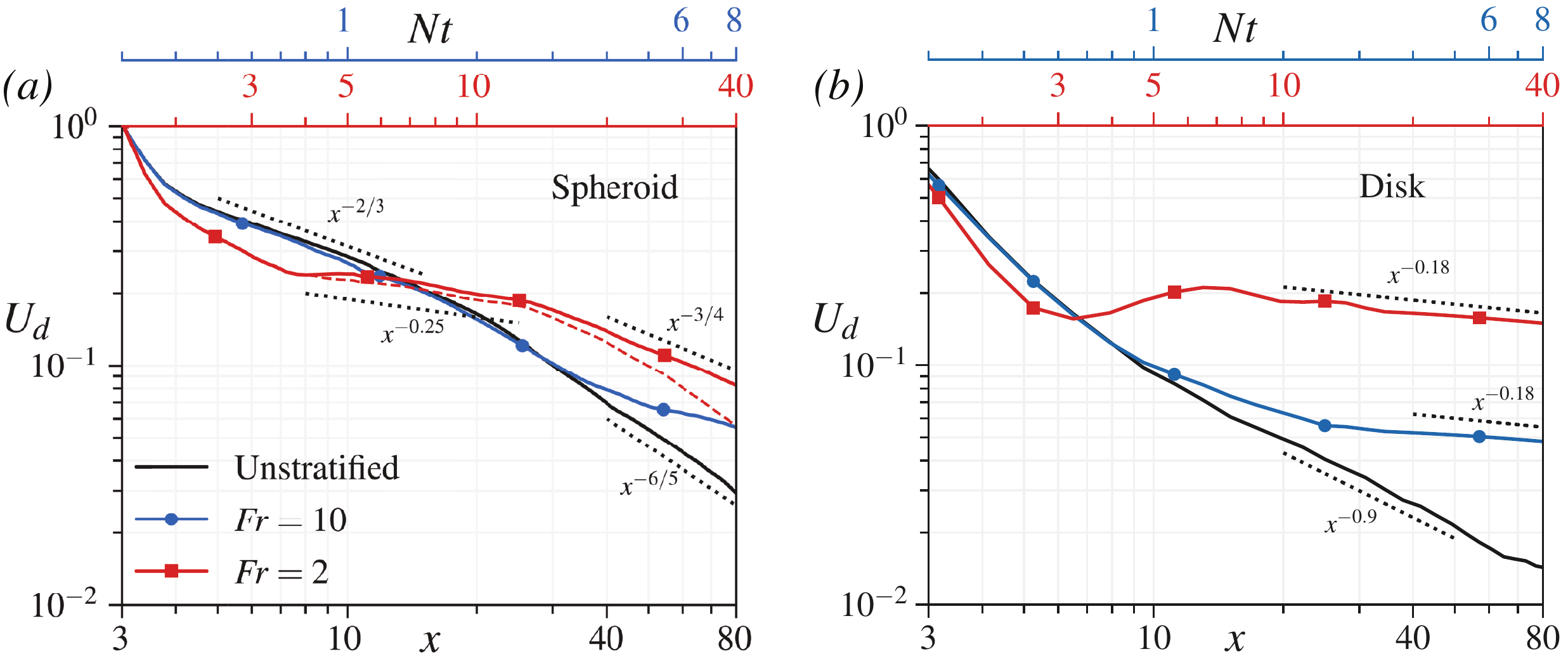}}
	\caption{Decay of the peak defect velocity in (a) spheroid and (b) disk. The red dashed line in (a) indicates the decay of the $\Fro = 2$ centerline defect velocity. For all other cases, centerline and maximum $U_d$ coincide. Note that the origin of the  $Nt$ scale is  1.5 for $\Fro = 2$ and is  0.3 for $\Fro = 10$.
	}
	\label{fig:ud_dis_spd}
\end{figure}

The decay rate of the mean defect velocity $U_d=U_\infty-U$ shows the different stages in the evolution of a wake. In a stratified environment, wakes transverse the 3D, NEQ and Q2D regimes \citep{Spedding1997}. Figure \ref{fig:ud_dis_spd} compares the decay of $U_d$ among the unstratified, $\Fro=10$, and $\Fro=2$ spheroid and disk wakes. To facilitate a one-to-one comparison, we present the disk data in the domain $3 \lesssim x \lesssim 80$, coinciding with the domain of the spheroid wake. Since $x$ is measured from the center of the body, the location of $x =3$ is in the near wake for the disk and is at the terminus of the body for the spheroid. The unstratified spheroid wake (figure \ref{fig:ud_dis_spd}a) shows a transition between the classical high-$\Rey$ decay $U_d\sim x^{-2/3}$ to $U_d\sim x^{-6/5}$ at $x \approx 20$ coinciding with the development of a helical structure (ONS21). The $\Fro = \infty$ disk wake decays as $U_d \sim x^{-0.9}$ from $10 \lesssim x \lesssim 65$ and transitions to the classical high-$\Rey$ decay of $x^{-2/3}$ afterward (CS20), as shown in figure \ref{fig:ud_dis_spd}(b). Compared to the disk, the $\Fro = 10$ and $\Fro=\infty$ spheroid wakes have a higher value of $U_d$, owing to  weaker near-wake entrainment and the slower  development of slender body wakes.

In the weakly stratified $\Fro=10$ regime, the defect velocity of the spheroid wake (figure \ref{fig:ud_dis_spd}a) evolves similarly to the unstratified wake until $Nt\approx 3.5$, when the decay rate slows down. However, at the same value of  $\Fro=10$ but for the disk wake (figure \ref{fig:ud_dis_spd}b), $U_d$ deviates from the unstratified case at $Nt \approx 1$. Based on $U_d$, the end of the 3D region and the beginning of the NEQ region of the spheroid wake occurs at $x\approx 30$ whereas in the disk it occurs at $x\approx 10$. We discuss the reason behind this delayed deviation of  the spheroid-wake $U_d$ from its unstratified counterpart  in \S 5.

At $\Fro = 2$, there are significant differences in $U_d$ evolution between the disk and spheroid wakes. In the $\Fro=2$ spheroid wake, $U_d$ shows an increased decay rate from the beginning. Although not shown here, the wake establishment is affected similarly to the $\Fro=1$ wake of the 4:1 spheroid of \cite{Ortiz-tarin2019}, where there was no 3D regime. The boundary layer evolution on the body and the separation are affected by stratification. 
At $Nt\approx \pi$, there is a sudden change in the decay rate due to the lee-wave-induced oscillatory modulation \citep{Pal2017} observed in the $3 \lesssim x \lesssim 10$ region. This oscillatory modulation gets weaker downstream as the lee-wave amplitude decreases with the distance from the source.

At $Nt \approx \pi$,  the wake transitions to the  NEQ stage where $U_d$ exhibits  a slower decay compared to both  the preceding stage and the following stage which commences at $Nt \approx 15$. Fitting a power law to the NEQ stage \revtwo{between $x=6-25$} results in a decay with $x^{-0.266}$. This decay is close to the -1/4 decay in the NEQ regime found in the experiments of \cite{Spedding1997} and later in numerical simulations \citep{Diamessis2011, Brucker2010,Redford2015,Pal2017}. \revtwo{More details about the fitting strategy can be found in ONS21.} At $Nt\approx 15$,  the spheroid wake transitions to the Q2D regime with a sharper decay and a power-law fit \revtwo{between $x=30-80$} results in  $U_d \sim x^{-0.72}$, which is close to the $x^{-0.75}$ behavior established by \cite{Spedding1997} for the Q2D regime. The $\Fro = 2$ disk wake shows a very different behavior. Until at least $x = 125$ ($Nt = 62.5$) -- the full extent of the computational domain -- the disk  wake exhibits no transition to the Q2D regime. Instead, after transitioning to the NEQ regime with a power law of  $U_d \sim x^{-0.18}$, the disk wake stays in that regime.
	
Thus, the NEQ regime in the spheroid wake at $\Fro = 2$ is significantly shortened compared to the disk wake, with it starting at $Nt \approx \pi$ and ending early at $Nt \approx 15$ when Q2D commences. In the experiments of \cite{Spedding1997}, the NEQ regime is reported to last until $Nt\approx 40$. Other temporal studies have found that the span of the NEQ regime depends on the Reynolds number. For example, in temporal simulations, \cite{Diamessis2011} found an increase of the NEQ duration to $Nt \approx 50$ when the Reynolds number increased to $\Rey=10^5$. \cite{Brucker2010} found a transition to a Q2D-type power law at $Nt\approx 100$. Only \cite{Redford2015} observed an earlier transition around $Nt \approx 25$. The reasons behind the early arrival of the Q2D regime in the spheroid $\Fro = 2$ wake will be discussed in \S 5.

\subsection{Evolution of the mean horizontal ($L_H$) and vertical ($L_V$) lengthscales} \label{sec:meanL}

\begin{figure}
	\centerline{\includegraphics[scale=0.5]{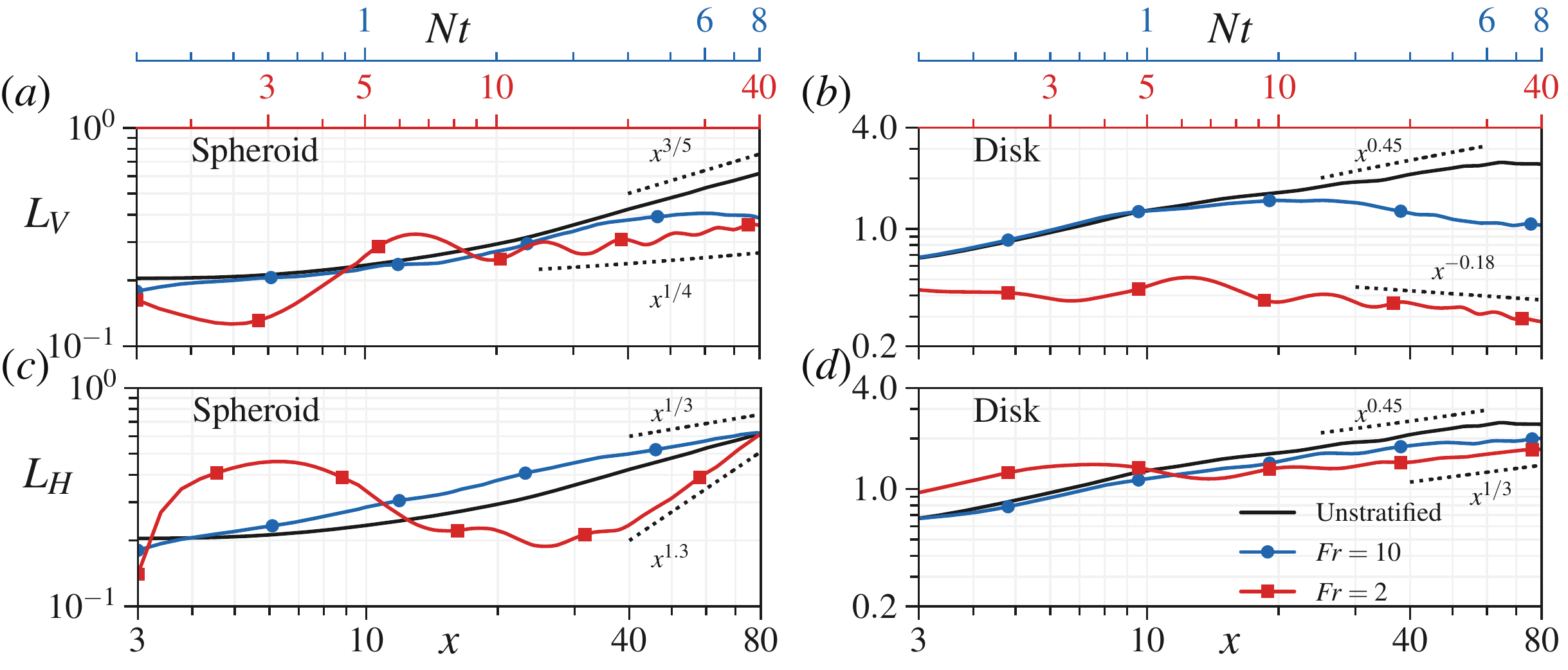}}
	\caption{Wake dimensions measured using the mean defect velocity $U_d$ for the spheroid (a,c) and disk (b,d) wakes in center-vertical (a,b) and center-horizontal (c,d) planes. The legends are same as in figure \ref{fig:ud_dis_spd}.}
	\label{fig:lmean_dis_spd}
\end{figure}

\begin{figure}
	\centerline{\includegraphics[scale=0.5]{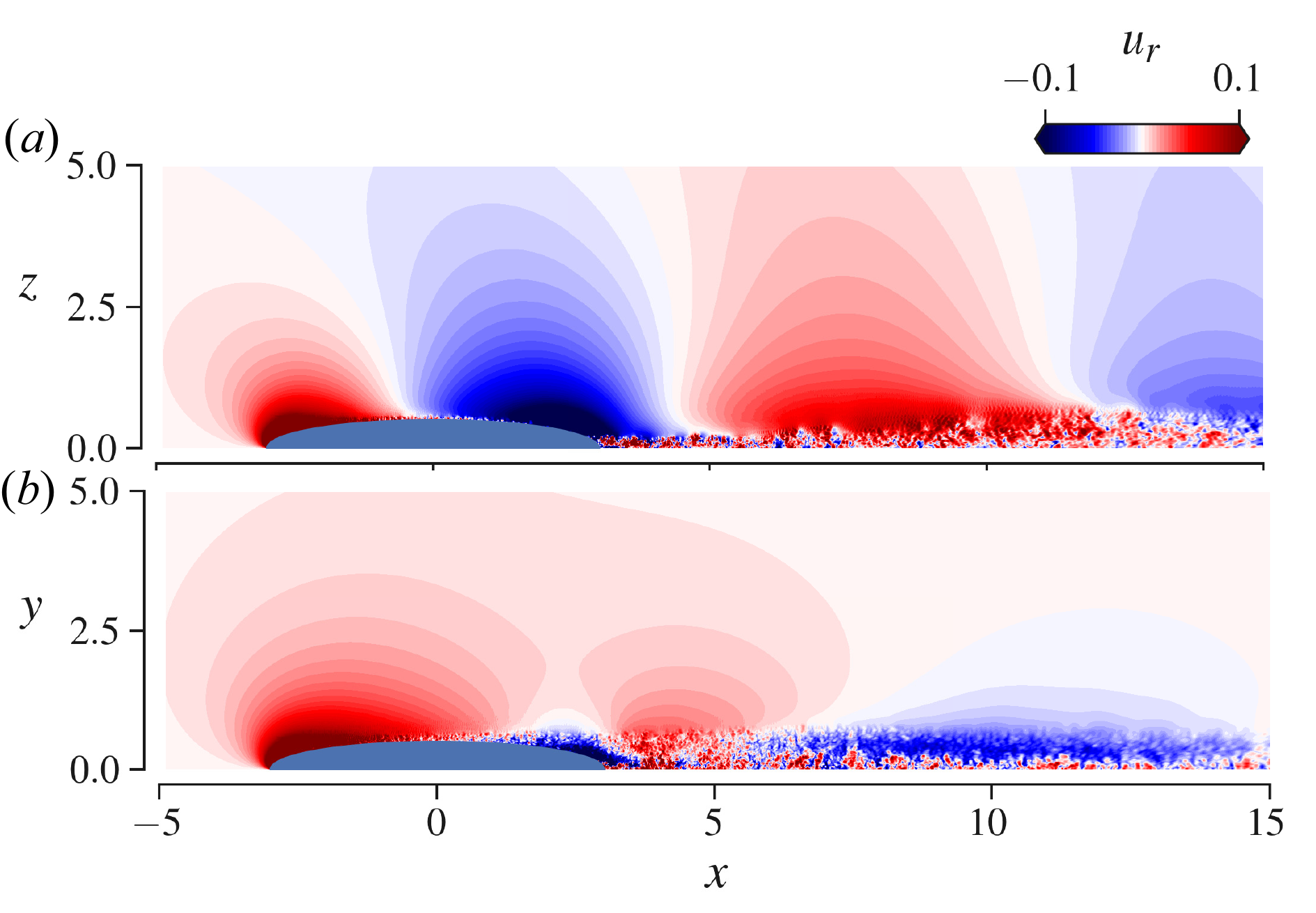}}
	\caption{Instantaneous radial velocity contours of the $\Fro=2$ spheroid wake in the (a) center-vertical and (b) center-horizontal planes.}
	\label{fig:igw_vis}
\end{figure}


The evolution of the mean wake dimensions in the spheroid and disk wakes is shown in figure \ref{fig:lmean_dis_spd}(a,c) and (b,d), respectively. $L$ is defined such that $U_\infty - U(L)=\frac{1}{2}U_d$. The subscripts $\{V,H\}$ indicate that these measures have been taken in the vertical and horizontal planes so that they represent the half height and the half width.

The wake of a slender body is generally thinner than its bluff body counterpart. Compared to the disk wake of CS20, the present unstratified wake is smaller by a factor of about 3 -- contrast black lines in figure \ref{fig:lmean_dis_spd}(a) to (b). The difference in wake size stems from the different near-wake features. Here, the initial non-dimensional wake width is around $0.2$ whereas its value for the disk is around 0.7. This observation agrees well with the scaling proposed by \cite{Tennekes1972} and used in stratified wake experiments by \cite{Meunier2004},  where the wake dimensions behind a body with diameter $D$ scale with the drag coefficient $\sqrt{C_D}$. We find that $C_D^{\mathrm{disk}} \approx 1.11$ and $C_D^{\mathrm{spheroid}} \approx 0.13$ resulting in $(C_D^{\mathrm{disk}}/C_D^{\mathrm{spheroid}})^{0.5} \approx 3.2$. Besides the initial dimensions, the near-wake growth rates  of the spheroid and disk are also very different. Whereas in the $x=3-20$ region the spheroid unstratified wake grows as $L \sim x^{0.2}$, the disk wake grows as $L \sim x^{0.45}$. Later, the growth rate of both wakes becomes comparable but the difference in size is already established and dictated by the near wake.

The evolution of the $\Fro=10$ spheroid wake height ($L_V$) is similar to its unstratified counterpart until $Nt\approx 3.5$ where the growth of $L_V$ slows down. While $L_V$ remains almost constant beyond $Nt\approx 4$, $L_H$ keeps increasing with a growth rate of $\sim x^{1/3}$. In the $\Fro = 10$ disk wake, the deviation from the $\Fro = \infty$ case happens at $x \approx 20$ ($Nt \approx 2$). Interestingly, after $Nt \approx 2$, the disk wake shows a continuous decrease in wake height. $L_H$ of both spheroid and disk wakes at $\Fro = 10$ closely follow the trend of the corresponding unstratified wake. See figures \ref{fig:lmean_dis_spd}(c,d).


The wake dimensions at $\Fro=2$ for both disk and spheroid show oscillations with a wavelength of $\lambda/D=2\pi \Fro$. This reveals the  influence of the steady lee waves especially on the wake height, see figure \ref{fig:lmean_dis_spd}(a,b). From $Nt=1-15$ the oscillations of $L_V$ and $L_H$ are consistent with the conservation of momentum deficit, i.e., to counteract the contraction of $L_V$ caused by buoyancy,  $L_H$ is enhanced. Note that these initial oscillations are of similar amplitude in both disk and spheroid. However, the relative change over the initial wake dimensions is much more pronounced in the spheroid wake ($\sim 10$ times) owing to its initial thinness. The influence of the lee waves on the spheroid $\Fro=2$ wake dimensions is illustrated by radial velocity contours in figure \ref{fig:igw_vis}. The wake width contracts significantly in the region where the vertical velocity of the lee wave induces a rapid increase of wake height. Starting at $Nt=20$, the width of the spheroid wake shows a rapid growth $L_H \sim x^{1.3}$ corresponding with: (i)  the development of the horizontal wavy motions observed in figure \ref{fig:fr2_wake_viz}(d) and (ii)  the arrival of the Q2D stage with $U_d \sim x^{-3/4}$. The growth of $L_V$ remains constant at a rate of $x^{0.25}$. Both, the very rapid growth of $L_H$ and the sustained increase in $L_V$ of the spheroid wake, are very different from the trends in the $\Fro = 2$ disk wake. In the disk wake, we find that, instead of an increase,  the vertical height exhibits a decrease ( $L_V \sim x^{-0.18}$)  at $x \gtrsim 20$. Furthermore, $L_H$ grows at $x^{1/3}$, a moderate rate relative to its rapid growth rate in the disk wake. 

%
%
		

\subsection{Comparison of flow topology between stratified spheroid and disk wakes}

\begin{figure}
	\centerline{\includegraphics[scale=0.4]{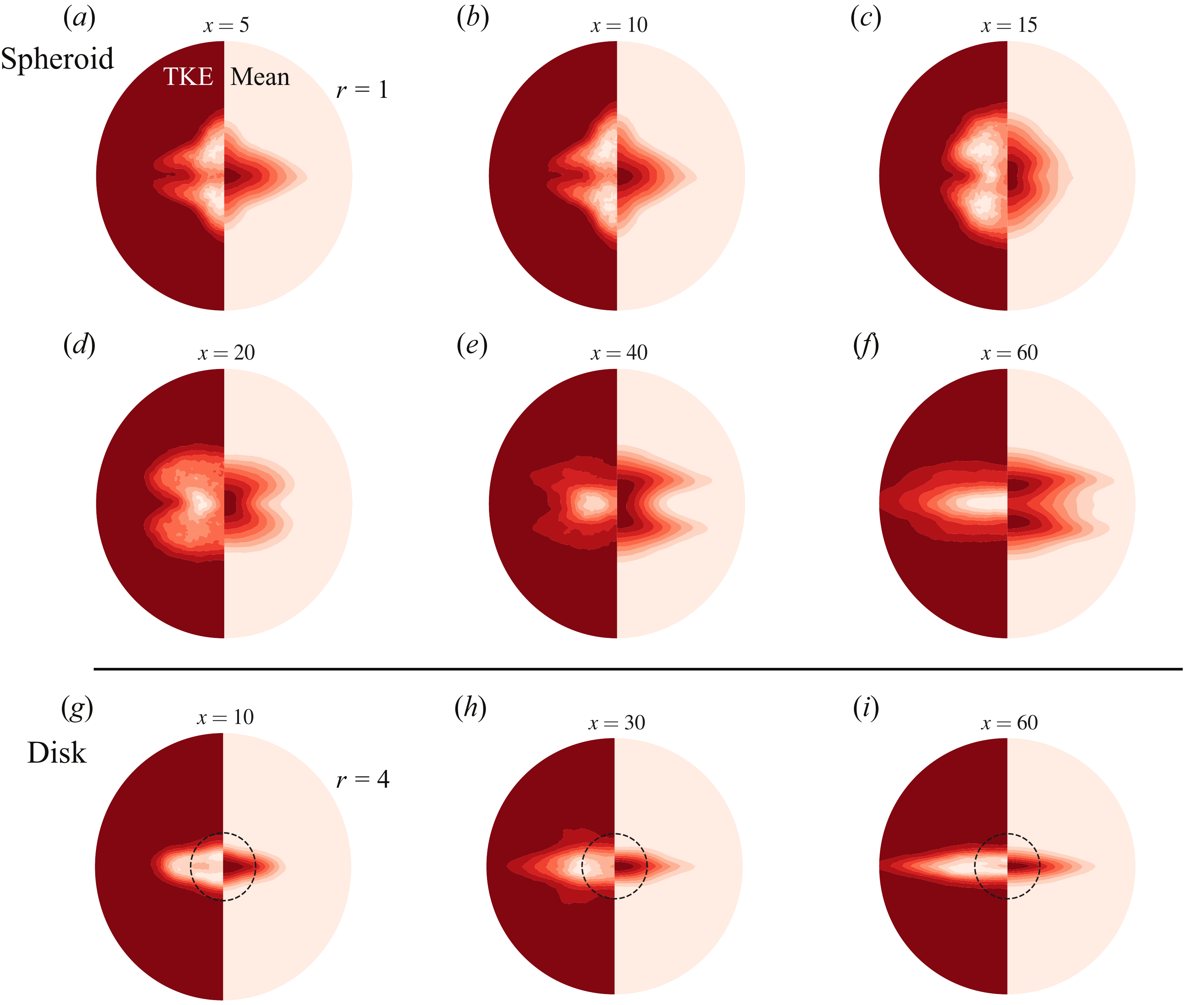}}
	\caption{$\Fro = 2$ wake of spheroid (a-f) and disk (g-i) at different streamwise locations $x$. Contours of mean streamwise velocity in the right half and TKE in the left half of each contour. Contour limits are between the minimum (red) and maximum (white) values of the respective quantity at a given $x$ with ten levels in between.  Radial extent span till $r = 1$ and $r=4$ for the spheroid and disk contours, respectively. The disk wake is larger than the spheroid wake as can be inferred from the $r =1$ circle  in the bottom row. }
	\label{fig:disk_spd_contours}
\end{figure}


The difference between the spheroid and disk  with regards to  the evolution of mean length scales  ($L_V$ and $L_H$), particularly at $\Fro=2$, points toward qualitative differences in the flow topology. To further characterize these differences in the $\Fro=2$ wake, figure \ref{fig:disk_spd_contours} shows contours of mean streamwise velocity ($U$)  at different streamwise locations for the  spheroid (top two rows)
 and the disk  (bottom row).
In each panel of figure \ref{fig:disk_spd_contours}, the right half shows $U_x$ and the left half shows turbulent kinetic energy (TKE), $E_T^{K}=(\langle u_{x}^{\prime 2} \rangle + \langle u_{y}^{\prime 2} \rangle + \langle u_{z}^{\prime 2} \rangle)/2$. 

 For the sake of brevity, we have not included contours of the $\Fro=10$ disk and spheroid wakes since their topology is similar -- the mean can be well approximated by a vertically-squeezed two-dimensional Gaussian while the TKE evolves from a bimodal (off-center peaks) distribution in the radial direction to a Gaussian at intermediate to late streamwise distances. In the case of the disk, the TKE evolves as a two-dimensional Gaussian right beyond the recirculation region.

We first discuss the disk wake (bottom row). The mean shows a monotonic spread in the horizontal direction and  resembles the shape of an ellipse or a two-dimensional Gaussian squeezed in the vertical direction. This shape does not change until the end of the computational domain at $x \approx 125$. The TKE for the disk wake also has a similar vertically squeezed appearance.


Turning to the spheroid wake, we find that its turbulence topology is different from that of the mean.  In the region $5 \leq x \leq 15$, TKE shows two off-center peaks reminiscent of the TBL shedding from a slender body \citep{Jimenez2010,Posa2016,Kumar2018,Ortiz2021} while mean $U_x$ shows a single central peak. At $x = 20$, we see the start of a horizontal contraction of the mean velocity in the central region of the wake and the emergence of a `butterfly' shape reminiscent of the separation and wake patterns observed in \cite{Ortiz-tarin2019}, where the $\Fro_c = 4/\pi \approx 1$ wake of a 4:1 spheroid was studied. Note that in this stage, the wake is thinner than taller, i.e.,  $L_H<L_V$. At $x \approx 20$, TKE starts transitioning from a bimodal distribution to a single peak near the center-horizontal plane. In the next section, we will analyze how the horizontal contraction of the mean wake  between $x = 20$ and  $x = 40$ results in an increased horizontal mean shear, resulting in the maximum TKE being produced close to the center-horizontal plane. This leads to a transition in the TKE topology from bimodal distribution to a squeezed-Gaussian  distribution at $x \gtrsim 40$. By $x \approx 60$, $U$ has been organized into two distinct layers while TKE is sustained between these two vertically off-center layers. Note that the multi-layered mean flow structure at late $x$ in the $\Fro = 2$ spheroid wake is reminiscent of the layered structure of the Q2D regime \cite{Spedding1997}.


Previously, temporal simulations \citep{Gourlay2001, Brucker2010, Redford2015} have shown that the mean and the turbulence can evolve differently. Indeed, the effect of buoyancy is `felt' very differently by the large and the small scales in the flow. The general trend is that, in the late wake, the turbulence occupies a smaller and smaller vertical fraction of the mean defect as time passes \citep{Redford2014}. Instead, the finding here for the spheroid wake  is the combined effect of having a wake in saturated-lee-wave state and initial off-center peaks of TKE, established by the TBL separation. These characteristics of the flow have not been not captured in temporal simulations since they have not accounted for  the wake generator and the steady lee waves. 

\section{Evolution of the turbulent flow in spheroid and disk wakes}

The energy of the flow is contrasted between spheroid and disk wake in this section. The turbulent kinetic energy (TKE also denoted as $E_K^{T}$), turbulent potential energy (TPE, $E_P^{T}$), mean kinetic energy (MKE, $E_K^{M}$) and mean potential energy (MPE, $E_P^{M}$) are defined as:

\begin{equation}
	E_K^{T}=(\langle u_{x}^{\prime 2} \rangle + \langle u_{y}^{\prime 2} \rangle + \langle u_{z}^{\prime 2} \rangle)/2\:, \qquad E_P^{T}=\gamma \langle \rho^{\prime} \rho^{\prime} \rangle /2,
\end{equation}

\begin{equation}
	E_K^{M}=(U_d^{2} + \langle u_{y} \rangle^{2} + \langle u_{z} \rangle^{2})/2\:,\qquad E_P^{M}=\gamma \langle \rho_{d} \rangle^{2} /2,
\end{equation}	
where $\gamma = g^{2}/\rho_o^{2}N^{2}$. In what follows,  trends of area-integrated values, denoted by $\{ \cdot \}$,  of these energy measures are reported. Area-integrated quantities are preferred because the peaks of mean and turbulence in stratified slender wakes are often times off-centered as can be seen in figure \ref{fig:disk_spd_contours}. The integration allows for a uniform comparison across cases. \revtwo{Also note that temporally averaged quantities are denoted by the angled brackets $\langle \cdot \rangle$}.

\subsection{Evolution of TKE, spectra and PE-to-KE ratios}

\begin{figure}
	\centerline{\includegraphics[scale=0.5]{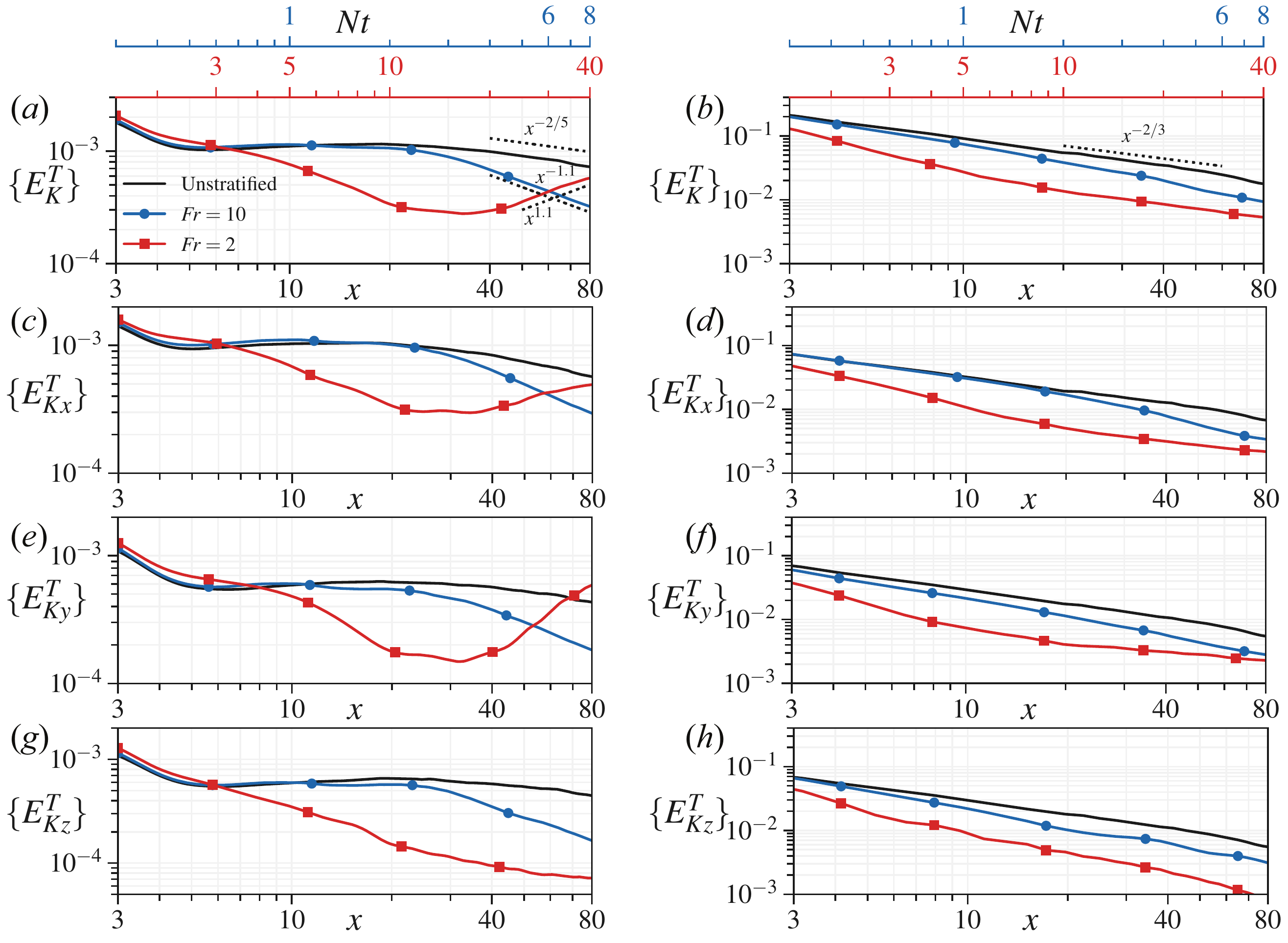}}
	\caption{Comparison of evolution of TKE between spheroid (left) and disk (right) wakes. (a,b) total TKE, (c,d) streamwise TKE, (e,f) spanwise TKE, and (g,h) vertical TKE.}
	\label{fig:energy_comps}
\end{figure}

The evolution of the area-integrated TKE of the disk and spheroid wakes is shown in figure \ref{fig:energy_comps}. 
The most noticeable aspect is that, for all $\Fro$ numbers, $\{E_K^{T}\}$ in spheroid wakes is an order of magnitude smaller than in corresponding disk wakes. Although it is not shown here, we found a similar result for the area-averaged TKE (over an area of $r=4D$ cross-section), instead of area-integrated TKE. 

The decay of the unstratified wake is studied in more detail in ONS21 for the spheroid  and CS20 for the disk.  In   unstratified flow, $\{E_K^T\}$ decays following $\{E_K^T\}\sim x^{-2/5}$ for the spheroid and the  disk wake follows $\{E_K^T\} \sim x^{-2/3}$. \revtwo{Note that these fits are empirical. Both decay rates are consistent with the decay of the peak TKE ($k$) and the growth of wake width ($L$) under the self-similarity framework, i.e., $\{E_K^{T}\} \sim k L^{2}$. Specifically, in the case of the spheroid $L\sim x^{3/5}$ and $k \sim x^{-8/5}$ \citep{Ortiz2021} and, in the case of the disk,  $L \sim x^{1/3}$ and $k \sim x^{-4/3}$ \citep{Chongsiripinyo2020}.}


At $\Fro=10$ and for both the spheroid and the disk, $\{E_K^{T}\}$ deviates from the unstratified case at \revtwo{between $Nt \approx 1$ and $ Nt \approx  3$}, see figure \ref{fig:energy_comps}(a,b).Interestingly, while the TKE is affected by stratification in the \revtwo{similar $Nt$ range} for both wake generators, the mean flow showed a different behavior. In the disk $\Fro=10$ wake, $U_d$ deviated from its unstratified counterpart at $Nt\approx 1$ while, in the spheroid, this change occurred later at $Nt \approx 3$. 
The onset of deviations from the unstratified case will be explained in more detail in the following subsections.
 


At $\Fro=2$, there is a striking influence of the wake generator on the evolution of TKE. Whereas in the disk wake,  the TKE decays monotonically, the far wake of the spheroid displays a rapid increase.
 
The disk wake shows a monotonic decay in $\{E_T^{K}\}$ and its individual components throughout $3 < x < 80$. Compared to the horizontal components, $\{E_{Tz}^{K}\}$ shows a sharper decay after $Nt \approx 10$ and turbulence anisotropy progressively increases. In the spheroid $\Fro=2$ wake, $\{E_T^{K}\}$ decays rapidly until $Nt\approx 10$.  However, after  $Nt\approx 10$, the decay slows down and is followed by a period of sustained growth starting at $Nt \approx 20$ and lasting until the end of the domain. The region of TKE growth coincides with the development of the large scale horizontal motions observed in figure \ref{fig:fr2_wake_viz}(d) and the rapid growth of $L_{H}$ shown in figure \ref{fig:lmean_dis_spd}(c). It also coincides with the accelerated decay rate of $U_d$ starting at $Nt\approx 20$.




Notice that right before the start of the rapid increase of TKE, the $\Fro=2$ wake has a configuration where $L_H<L_V$ (figure \ref{fig:disk_spd_contours}d). 
The horizontal response of the flow to the strong lee waves is what sustains this configuration. It is only after their strength subsides that the control on the wake is released to allow the horizontal wavy motion to develop. The rapid development of this motion coincides with the rapid increase in horizontal TKE, namely $\{E^T_{Ky}\}$ and $\{E^T_{Kx}\}$  as seen in figure \ref{fig:energy_comps}(c,e). To the best of the authors' knowledge this is the first wake study, resolving the flow at the body, that observes an increase of fluctuation energy with downstream distance instead of its usual decrease. 

\begin{figure}
	\centerline{\includegraphics[scale=0.5]{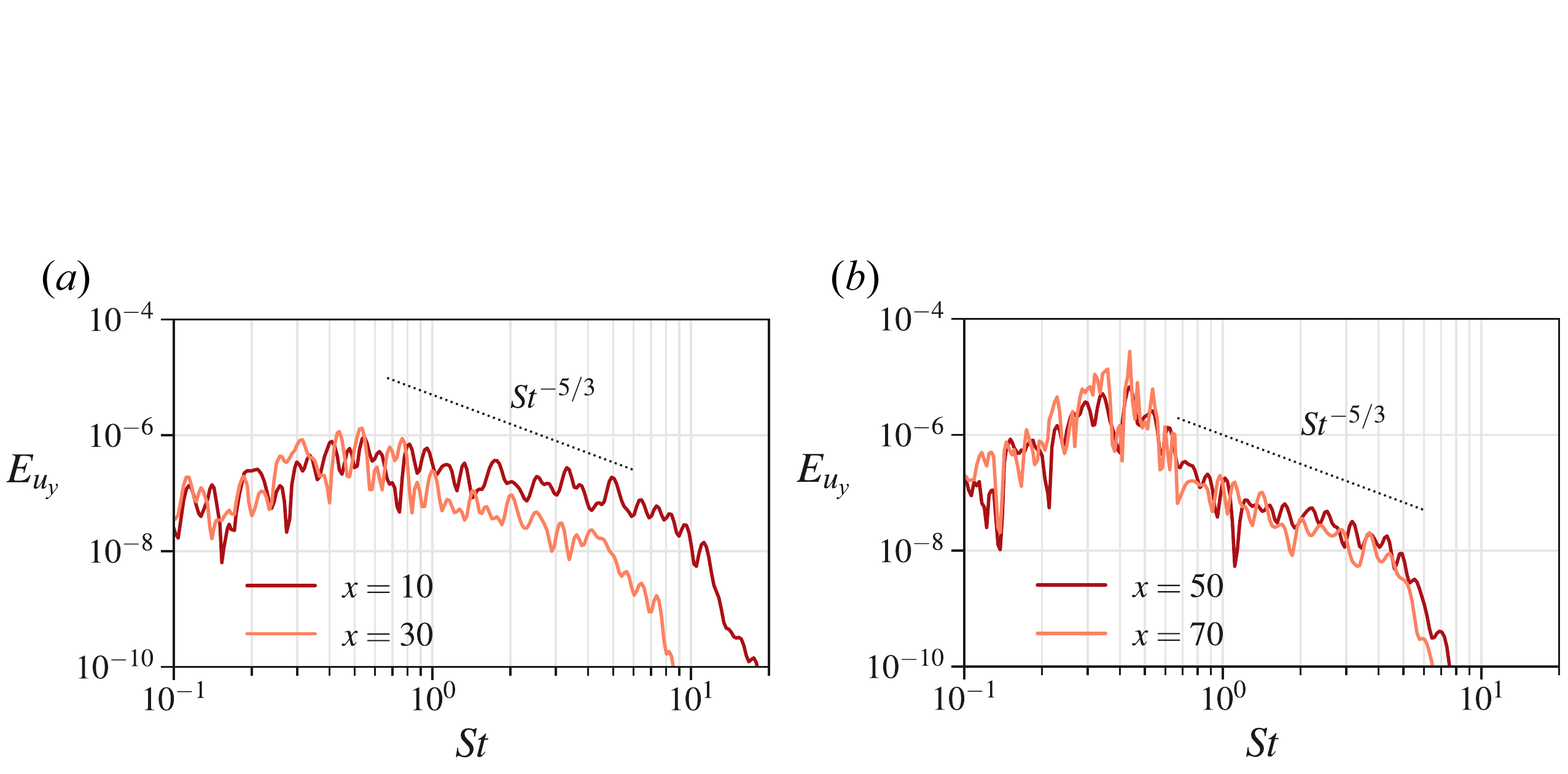}}
	\caption{Energy spectra of the $\Fro = 2$ spheroid wake computed with the spanwise velocity fluctuations at the centerline at (a) $x = 10, 30$ and (b) $x = 50, 70$.}
	\label{fig:SpectraFr2}
\end{figure}
 
To further quantify the horizontal wavy motions observed in figure \ref{fig:fr2_wake_viz}(d),  the energy spectra of the spanwise velocity are computed at the centerline. These spectra are compared between locations before (figure \ref{fig:SpectraFr2}a)  and after (figure \ref{fig:SpectraFr2}b) the start of the the TKE increase associated with the Q2D regime. The spectra before $x < 40$ do not show preferential energization of  the low frequencies. This finding is consistent with the visualizations in figure \ref{fig:fr2_wake_viz} where the intermediate wake does not show any sign of large scale motions. Beyond $x = 40$, however, spectra show a strong peak at Strouhal number \revtwo{$\Str=fD/U_{\infty}$} $\approx 0.35$ (figure \ref{fig:SpectraFr2}b). This value of Strouhal number agrees with the approximate wavelength of structures in figure \ref{fig:fr2_wake_viz}(d), where the wavelength $\lambda/D \approx 1/\Str$. To summarize, the arrival of the Q2D regime in the $\Fro = 2$ wake is accompanied by a strong increase in TKE (figure \ref{fig:energy_comps}a) and the appearance of large scale motions in the center-horizontal plane (figure \ref{fig:fr2_wake_viz}d and \ref{fig:SpectraFr2}b).

In the $\Fro = 2$ disk wake,  the vortex shedding mode at $\Str = fD/U_\infty \approx 0.13-0.14$ is dominant throughout the whole domain. The horizontal meanders, which are prevalent in the spheroid wake, are absent in the disk wake  at least until the end of the domain at $x/D = 125$. Since the vortex shedding mode, its long-lasting effect on the wake, and its internal wave field are described in detail by \cite{nidhan2022analysis},  we do not discuss these aspects further.


\begin{figure}
	\centerline{\includegraphics[scale=0.5]{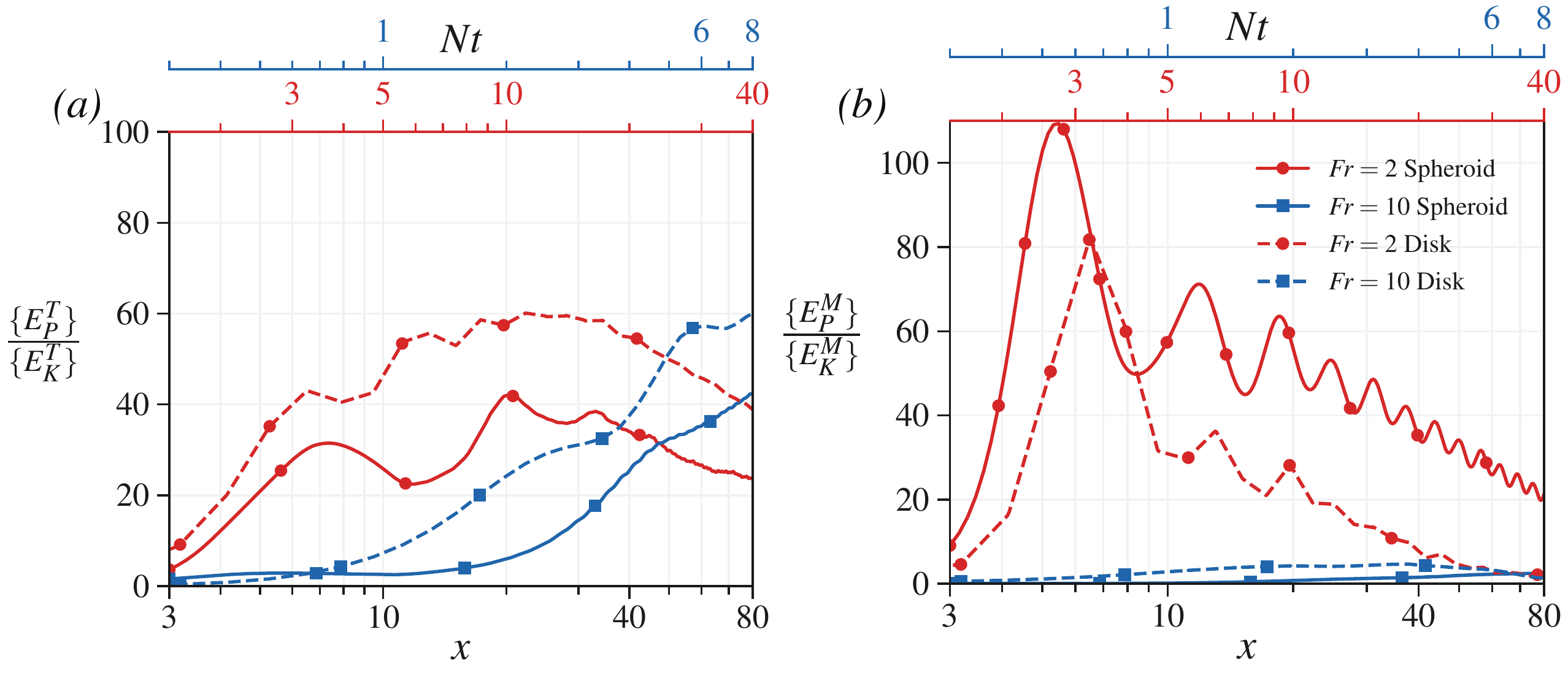}}
	\caption{Ratio of area-integrated (a) turbulent potential energy to turbulent kinetic energy and (b) mean potential energy to mean kinetic energy in stratified spheroid and disk wakes.}
	\label{fig:energy_ratio}
\end{figure}

The spheroid wake has significantly lower TKE content relative to the disk wake and also a different distribution of the mean momentum. The content of potential energy relative to that of kinetic energy is also of interest. Figure \ref{fig:energy_ratio} shows the ratio of area-integrated potential energy (PE) to kinetic energy (KE). Both fluctuating (\ref{fig:energy_ratio}a) and mean (figure \ref{fig:energy_ratio}b)  components are shown at $\Fro = 2$ and $10$ for both wake generators. 

At $\Fro = 10$, the turbulent PE-to-KE ratio increases steadily in both disk and spheroid cases indicating an increasing influence of buoyancy on turbulence (figure \ref{fig:energy_ratio}a). In the $\Fro = 2$ wakes, the turbulent PE-to-KE ratio peak around $x \approx 30$ in both cases and decay afterward. By the end of the measurement region at $x = 80$, the turbulent PE-to-KE ratios are similar across $\Fro = 2$ and $10$ wakes.  Thus, stratification and body shape do not qualitative affect   the ability of turbulence to stir the density field in the intermediate and far wake.
 Quantitatively, the TPE-to-TKE ratios are somewhat higher for the disk relative to the spheroid for both wakes.
	
The mean PE-to-KE ratios in the $\Fro = 10$ wakes (figure \ref{fig:energy_ratio}b) are minuscule compared to their turbulent counterparts. At $\Fro = 2$, the mean-based  ratio is much larger, particularly close to the wake generators, pointing towards a strong influence of the steady lee waves. Both disk and spheroid mean-based ratios oscillate with a characteristic lengthscale corresponding to the wavelength of steady lee waves at $\Fro = 2$. It is particularly revealing as to how much larger is the magnitude of $\{E^M_P\}/\{E^M_K\}$ in the spheroid wake compared to the disk, as it explains why the spheroid flow is so strongly modulated by the lee waves. It is worth noting that comparison of the absolute value of MPE between the  disk  and the spheroid reveals that it is the disk that has the larger MPE,  an order of magnitude larger. The amplitude of the lee wave generated by the disk is  larger than that of the spheroid by about a factor of 2.

\subsection{Analyses of the spheroid TKE budget terms}

To quantitatively understand the origin of TKE increase in the $\Fro = 2$ spheroid wake, we look into the different terms of the TKE transport equation: 

\begin{equation}
\label{eq:turb_budget}
U_i\frac{\partial{E^T_K}}{\partial{x_i}}+\frac{\partial{T_i}}{\partial{x_i}}=P-\varepsilon+B\:,
\end{equation}
where $P$ is the turbulent production, $\varepsilon$ is the turbulent dissipation and $B$ denotes the turbulent buoyancy flux. These quantities are defined by:

\begin{equation}
\label{eq:dissturb}
P=-\langle u'_iu'_j \rangle \frac{\partial U_i}{\partial x_j},\quad
\varepsilon = 2\nu\langle s'_{ij}s'_{ij}\rangle -\langle\tau'^{s}_{ij}s'_{ij}\rangle \;,\quad B=-\frac{g}{\rho_0}\langle \rho' u_z'\rangle\;,
\end{equation}

where $s_{ij}=(\partial_j u_i+\partial_i u_j) /2$ is the strain-rate tensor and $\tau^s_{ij}=-2\nu_ss_{ij}$ is the subgrid stress tensor. The contribution of the subgrid term to the TKE transport equation is found to be small.

The turbulent production is a source of TKE and a sink in the MKE equation. The turbulent buoyancy flux transfers energy between TKE and TPE and the turbulent dissipation is a sink of TKE. Along with the sinks and sources of energy, there is a term responsible for the spatial redistribution of TKE, the turbulent transport,

\begin{equation}
T_i = \frac{1}{2}\langle{u'_{i}u'_{j}u'_{j}}\rangle+\langle u'_{i}p'\rangle-2\nu\langle{u'_js'_{ij}}\rangle-\langle{u'_j\tau'^s_{ij}}\rangle\;.
\end{equation}
The turbulent transport  redistributes energy, primarily in the $y$-$z$ plane, and its contribution to the area-integrated budget is negligible. 

\revone{The production term in equation (\ref{eq:turb_budget}) can be further expanded as follows:

\begin{equation}
	\label{eq:production_terms}
	P=-\langle u'_xu'_j \rangle \frac{\partial U_x}{\partial x_j} -\langle u'_yu'_j \rangle \frac{\partial U_y}{\partial x_j} -\langle u'_zu'_j \rangle \frac{\partial U_z}{\partial x_j}.
\end{equation}

For the stratified wakes at hand, $U_y, U_z << U_x$. Hence equation (\ref{eq:production_terms}) can be further simplified as below:

\begin{equation}
	\label{eq:production_terms_simplified}
	P=-\langle u'_xu'_x \rangle \frac{\partial U_x}{\partial x} -\langle u'_xu'_y \rangle \frac{\partial U_x}{\partial y} -\langle u'_xu'_z \rangle \frac{\partial U_x}{\partial z}.
\end{equation}

In a wake developing in the $x$ direction, $P$ is primarily dominated by the transverse shear terms since $|\partial U_x/\partial y|, |\partial U_x/\partial z| >> |\partial U_x/\partial x|$. Hence in the discussions on $\{P\}$ to follow, we focus on the transverse terms, namely $P_{xy} = \langle u'_x u'_y \rangle\partial_y U_x $ and $P_{xz}=\langle u'_x u'_z\rangle\partial_z U_x$.
}

\begin{figure}
	\centerline{\includegraphics[scale=0.5]{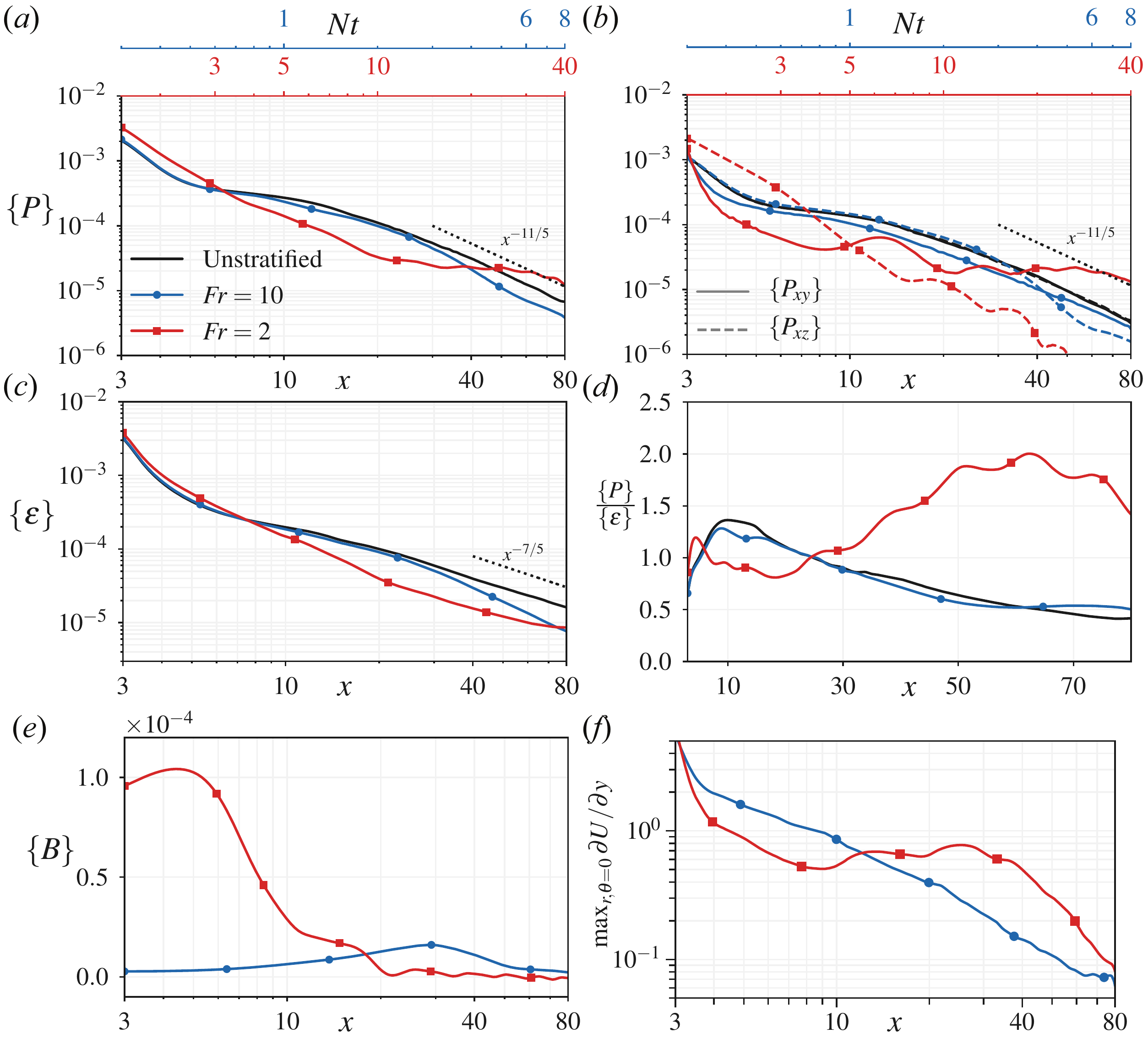}}
	\caption{Spheroid wakes. (a) Area-integrated production. (b) Main components of the turbulent production. (c) Area-integrated dissipation. (d) Ratio between area-integrated production and dissipation. (e) Area-integrated buoyancy flux. (f) Maximum value of mean horizontal shear $\partial U_x/\partial y$.}
	\label{fig:tke_budget_spd}
\end{figure}

Figure \ref{fig:tke_budget_spd} shows the evolution of the area-integrated terms in the TKE transport equation of the spheroid wake. We do not present the TKE transport terms of the disk wakes here as they are presented and discussed in great detail in CS20. \revone{We first discuss the trends of the area-integrated production, $\{P\}$, and its components in $\Fro = 2$ and $10$ spheroid wakes. This is followed by a discussion of the area-integrated dissipation, $\{\varepsilon\}$. Thereafter, the variation of $\{P\}/\{\varepsilon\}$ sheds light on why an increase in TKE is observed in the $\Fro=2$ spheroid wake for $x>30$.}

$\{P\}$ in the $\Fro=10$ spheroid wake decays at a rate comparable to its unstratified counterpart until $Nt \approx 3$ when it starts to deviate (figure \ref{fig:tke_budget_spd}a). $U_d$ in the $\Fro = 10$ spheroid wake also starts deviating around $Nt \approx 3$ (figure \ref{fig:ud_dis_spd}a). At $Nt \approx 3$, the contribution of $P_{xz}=\langle u'_x u'_z\rangle\partial_z U_x$ starts declining due to a reduction in $\langle u'_x u'_z \rangle$ {(not shown here for brevity)}. At $Nt \approx 4.5$, the contribution of the lateral  production $P_{xy}$ exceeds that of the vertical production $P_{xz}$. 

\revone{The reduction of $\{P_{xz}\}$ in the $\Fro=10$ wake coincides with the transition of $U_d$ to a slower decay rate. In figure \ref{fig:tke_budget_spd}(b), we note that $\{P_{xz}\}$ starts decaying faster than its unstratified counterpart between $3 < Nt < 4$ (or $30 < x < 40$). In figure \ref{fig:ud_dis_spd}(a), we see that this the location where $U_d$ of the $\Fro = 10$ spheroid wake starts deviating from the $\Fro = \infty$ spheroid wake. At the same location that $\{P_{xz}\}$ starts decaying rapidly, there is a maximum in the buoyancy flux $\{B\}$ (figure \ref{fig:tke_budget_spd}e) and a slowdown in the growth of $L_V$ (figure \ref{fig:lmean_dis_spd}a). Buoyancy is starting to affect the wake decay.}


The initial value of $\{P\}$ in the $\Fro=2$ wake is higher than in the unstratified case. The distinct separation pattern and the vertical contraction of the mean flow (figure \ref{fig:lmean_dis_spd}a) lead to an increase in the vertical shear and hence $\{P_{xz}\}$, see red dashed lines in figure \ref{fig:tke_budget_spd}(b). This initially high value of $\{P_{xz}\}$ rapidly decays as $L_V$ increases, leading to a reduction in the vertical shear. Note that the beginning of the $U_d$-based NEQ regime at $Nt\approx \pi$ also coincides with this reduction of $\{P_{xz}\}$. However, the mechanism at $\Fro=2$ is different from that of the $\Fro=10$ wake. Whereas at $\Fro=10$, the reduction of $\langle u'_x u'_z \rangle$ causes the decay of $\{P_{xz}\}$, at $\Fro=2$ the decay of $\langle u'_x u'_z \rangle$ is accompanied by a sudden reduction of $\partial_z U_x$ (not shown here). The reduction in vertical shear is driven by the expansion of $L_V$ due to lee wave modulation in figure \ref{fig:lmean_dis_spd}(a).

As the modulation of the wake by the lee waves continues, the horizontal production is enhanced by the strong reduction in $L_H$ (figure \ref{fig:lmean_dis_spd}c) in the NEQ region during $7 < x < 20$. The value of $\{P_{xy}\}$ overtakes $\{P_{xz}\}$ at $Nt\approx 5$. 
The magnitude of $\{P_{xy}\}$ remains nearly constant until the end of the domain. Figure \ref{fig:tke_budget_spd}(f) shows the maximum of the horizontal mean shear ($\partial U_x/\partial y$), i.e., $\mathrm{max}_{(r,\theta=0)} \partial U_x/\partial y$, in the central streamwise-horizontal plane for both wakes. The mean shear  in the  $\Fro = 2$ wake increases in the region $10 < x < 30$, exactly in the region where the TKE decay starts plateauing (see figure \ref{fig:energy_comps}a). The enhanced mean horizontal shear at $\Fro =2$  prevents the horizontal production from decaying monotonically unlike $\Fro = 10$.

\revone{Figure \ref{fig:tke_budget_spd}(c) shows the evolution of $\{\varepsilon\}$ as a function of $x$ for $\Fro = \infty, 10$ and $2$ spheroid wakes. In the $\Fro = 10$ case, the decay of $\varepsilon$ is similar to that of the unstratified wake until $Nt\approx 3$ after which the decay rate increases slightly. The $\Fro = 2$ wake dissipation shows a sharper decay until $x \approx 20$. After $x \approx 20$, the decay rate appears to be closer to the $\Fro =\infty$ decay rate of $x^{-7/5}$ \citep{Ortiz2021}.

Since the turbulent dissipation keeps decreasing monotonically in the $\Fro = 2$ spheroid wake (figure \ref{fig:tke_budget_spd}c), while the production tends to asymptote for $x>30$ (figure \ref{fig:tke_budget_spd}a), the value of $\{P\}/\{ \varepsilon\}$ becomes greater than $1$ in the $\Fro=2$ spheroid wake, see figure \ref{fig:tke_budget_spd}(d), explaining the rapid increase of TKE beyond $x \approx 30$. Note that the $\{P\}/\{ \varepsilon\}$ ratio oscillates with the characteristic wavelength of the lee waves revealing the strong influence of wave-related buoyancy effects on the energetics of the $\Fro=2$ wake.}

One of the features of the arrival of the NEQ regime reported in previous studies is the radiation of internal gravity waves \citep{Abdilghanie2013,DeStadler2012,Rowe2020}. 
In these spheroid wakes, we find that the integrated wave flux remains negligible compared to the other terms in the TKE budget and hence is not shown here. The small magnitude of the turbulent wave flux is consistent with the findings of \cite{Meunier2018} who found that the magnitude of the wake-generated waves depends on the body drag coefficient --  and the 6:1 spheroid has a very low drag compared to bluff bodies.



\subsection{Early arrival of the Q2D regime in the $\Fro = 2$ spheroid wake}

\revtwo{Remember that the Q2D regime in stratified wakes is characterized by a faster decay of the mean defect velocity ($U_d \sim x^{-3/4}$) and the organization of wake into distinct layers in the vertical direction. Due to the strong effects of buoyancy, the fluid motions are primarily confined to the horizontal plane. However, the vertical variations in the velocity profiles still exist during the Q2D regime, primarily in the form of layered structures.} 

A key difference between the $\Fro = 2$ disk and spheroid wakes is the early arrival of the Q2D regime in the spheroid wake. Whereas in the disk wake, CS20 did not observe a transition to the Q2D in a domain that extended up to $x = 125$ ($Nt = 62.5$), here we observe a transition at $x \approx 40$ ($Nt \approx 20$).  The early transition in the spheroid wake is a consequence of the strong modulation of the intermediate wake by buoyancy, an  effect that occurs for bodies with large aspect ratio ($L/D$) and, specifically, when $\Fro$ is near its critical value $\Fro_c = L/D\pi$. 
	
Figure \ref{fig:lmean_dis_spd}(c) shows that $L_H$ in the $\Fro = 2$ spheroid wake contracts in the region $5 < x < 30$ as a response to the expansion of $L_V$ (figure \ref{fig:lmean_dis_spd}a) by steady lee waves. This phenomenon leads to the `butterfly' shaped structure of mean $U_x$ (figure \ref{fig:disk_spd_contours}(d,e)) which was also observed at a lower Reynolds number, $\Rey=10^4$ in \cite{Ortiz-tarin2019}. In that study, which was performed at $\Rey = 10^4$, the spheroid wake at critical $\Fro_c$ relaminarized. Here, at a higher $\Rey = 10^{5}$, the flow response at the resonant state is quite different. The constriction of $L_H$ leads to an enhancement in the mean horizontal shear shown in figure \ref{fig:tke_budget_spd}(f). This enhancement significantly slows down the decay of horizontal production (figure \ref{fig:tke_budget_spd}b). While $\varepsilon$ continues to decay, $\{P\}/\{\varepsilon\}$ becomes $>1$ leading to an increase of TKE for $x >40$ (figure \ref{fig:energy_comps}a). In figure \ref{fig:disk_spd_contours}(d-f), we also see the maximum TKE location moving to the wake axis  from its off-center location in the near wake. It is worth noting that the enhanced $\{P\}$ that acts as a source of TKE is a sink for the mean energy. The sharp increase in TKE leads to a faster decay of $U_d$, and the $\Fro =2$ wake transitions to the Q2D regime early on, at $x \approx 40$. 

In the disk wake, $L_H$ is initially  $3-4$ times larger than in the spheroid wake. While the lee wave modulation of $L_H$ is present in the disk wake as well (figure \ref{fig:lmean_dis_spd}d), its amplitude relative the original value of $L_H$ is quite small. Hence the horizontal mean shear (not shown here) and the horizontal production in the $\Fro = 2$ disk wake continue to decay unlike in the spheroid wake.

The arrival of the Q2D regime in the spheroid wake is also accompanied by distinctive features of the Q2D regime reported in  previous literature. Figure \ref{fig:fr2_wake_viz}(d) shows lateral meanders in the late $\Fro = 2$ wake similar to the lateral meanders in temporal simulations in the literature \citep{Brucker2010,Diamessis2011}, albeit the temporal-simulation meanders occur much later in $Nt$ units. As noted during the discussion of spectra, the waviness in the late wake has a characteristic frequency $\Str \approx 0.35$. The mean wake in the Q2D regime has a layered topology (figure \ref{fig:disk_spd_contours}f) as reported by \cite{Spedding2002} and \cite{Chongsiripinyo2017}. The turbulence state in the Q2D regime is characterized by weak vertical fluctuations $u_z' << u'_h$ \citep{Spedding1997} with $\{E_{Kz}^{T}\}/\{E_{Kh}^{T}\} < 0.1$ at $x > 60$ -- where subscript $h$ denotes the horizontal component of the fluctuations. Since the Q2D regime of the spheroid wake is in a relatively early phase,  pancake vortices do not appear until the end of the simulation domain, $ x = 80$.

\subsection{Late transition to the NEQ regime in the $\Fro = 10$ spheroid wake}

\begin{figure}
	\centerline{\includegraphics[scale=0.5]{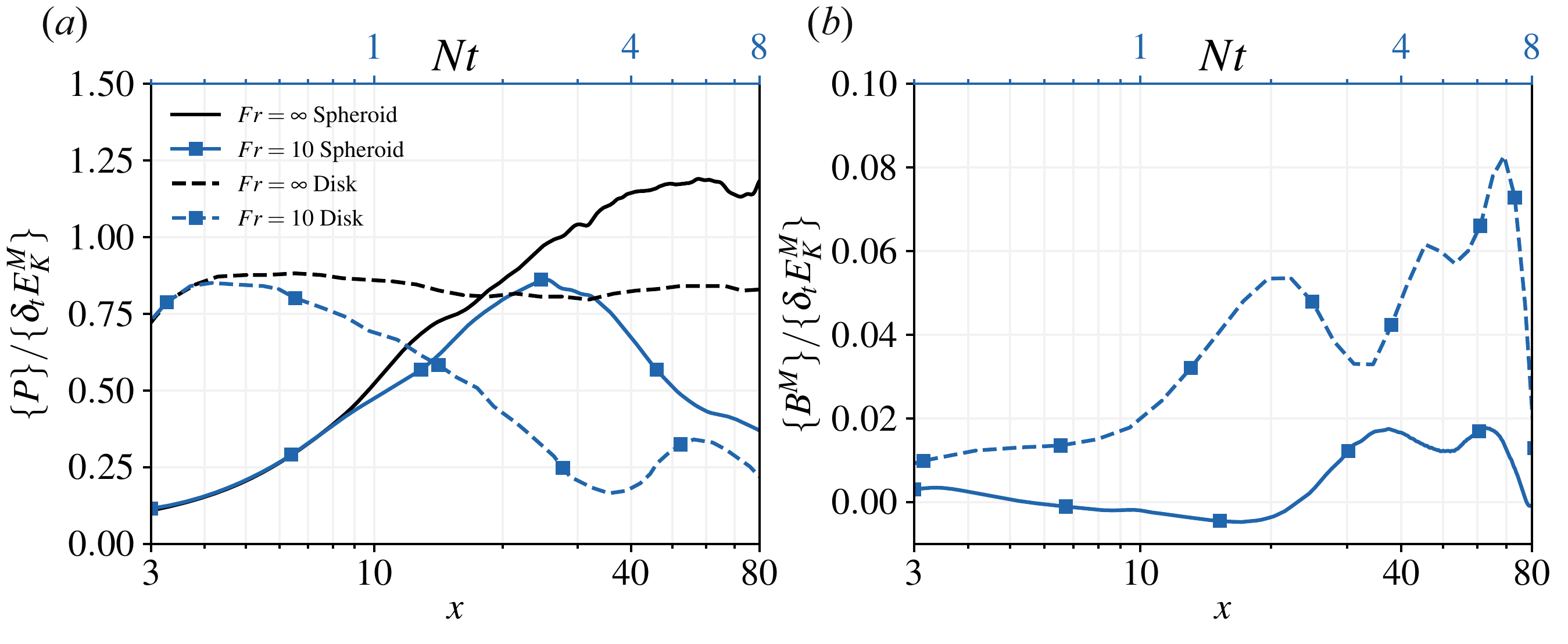}}
	\caption{Area-integrated (a) production and (b) buoyancy flux in the disk and spheroid $\Fro = 10$ wakes. The unstratified wake ($\Fro = \infty$) production is also shown in (a). The terms are normalized by the Lagrangian rate of change of their corresponding mean kinetic energy $\{\delta_t E_K^{M}\} = \{E^{M}_K\}U_\infty/x$.}
	\label{fig:lagrangian_budget_terms_Fr10}
\end{figure}

At $\Fro=10$, the main difference between the disk and spheroid wakes is the location at which $U_d$ deviates from the unstratified counterpart, i.e., the transition point to the NEQ regime. In the spheroid wake this transition occurs around $Nt\approx 3$, whereas in the disk it occurs at $Nt\approx 1$, compare the $\Fro = 10$ curves between figure \ref{fig:ud_dis_spd} (a) and (b). \revtwo{The beginning of the NEQ regime is marked by a slowdown in the decay rate of $U_d$ in the stratified wake as compared to its unstratified counterpart. The NEQ regime is also characterized by the progressively increasing
 anisotropy between the horizontal and vertical components of TKE.}

To understand better how this transition occurs we  look into the mean kinetic energy (MKE) transport equation. \revtwo{The TKE transport equation was given in equation (\ref{eq:turb_budget}). In a similar fashion, the MKE transport equation is given by:

\begin{equation}
	\label{eq:mean_budget}
	U_i\frac{\partial{E^M_K}}{\partial{x_i}}+\frac{\partial{T^M_i}}{\partial{x_i}}=-P-\varepsilon^M+B^M\:,
\end{equation}
where superscript $M$  denotes the mean counterparts of  terms in equation (\ref{eq:turb_budget}).
} 

\revtwo{Figure \ref{fig:lagrangian_budget_terms_Fr10}(a) shows the area-integrated production $\{P\}$ normalized by the Lagrangian change of the MKE ($\{E_K^{M}\}U_{\infty}/x$) in the disk and spheroid wakes at $\Fro = \infty$ and $10$. Likewise, figure \ref{fig:lagrangian_budget_terms_Fr10}(b) shows the normalized area-integrated mean buoyancy flux $\{B^{M}\}$.
Normalization by the Lagrangian change of MKE allows us to quantify the individual importance of each budget term to the change of MKE . 

Broadly,  the mean and  turbulence quantities in stratified wakes deviate from their unstratified counterparts as a result of buoyancy. In the MKE transport equation, buoyancy can  affect (a) the turbulence production \revtwo{$P$} as an indirect effect and (b) the MKE $\leftrightarrow$ MPE transfer, \revtwo{through the mean buoyancy flux $B^M$}. In both the disk and spheroid $\Fro = 10$ wakes, we find that the contribution of the mean buoyancy flux to the MKE transport is significantly smaller than the contribution of the production, particularly for $x \lessapprox 40$. Now turning to the area-integrated production in figure \ref{fig:lagrangian_budget_terms_Fr10}(a), we find that the production at $\Fro = 10$ displays a strong reduction from the $\Fro = \infty$ wake later in the spheroid at $x \approx 30$ ($Nt \approx 3$) compared to the disk {where a similar strong reduction occurs at $x \approx 10$} ($Nt \approx 1$). This explains  the late transition of the spheroid wake  to the NEQ regime, i.e.  $U_d$ deviates from unstratified behavior later ($Nt \approx 3$) for the spheroid than the $Nt \approx 1$ transition point for the disk.

The decreased production in the $\Fro=10$ wakes of both bodies is a consequence of the decreased $\langle u'_xu'_z\rangle$ correlation (not shown here for brevity) in stratified turbulent shear flows~\citep{JacobitzSV:1997}. In stratified wakes, this buoyancy effect has been observed experimentally by \cite{Spedding2002} and numerically by \cite{Brucker2010}. 
 The deviation between production of the $\Fro = 10$ and $\infty$ wakes (figure \ref{fig:tke_budget_spd}a) is indeed caused by a reduction in its $\{P_{xz}\}$ component, see blue dashed lines in figure \ref{fig:tke_budget_spd}(b). 
}

\section{Evolution of the local flow state and its trajectory in the phase space
}


In previous sections, we showed how the spheroid and the disk wake do not transition between the 3D, NEQ and Q2D regimes at the same $Nt$. In this section, we examine the evolution of  key {local} non-dimensional numbers describing the mean and fluctuating state to  explore their roles. These non-dimensional numbers are local, streamwise-varying measures of stratification (Froude number) and the dynamical range of inertially dominated scales (Reynolds number). We also plot the trajectory of each wake in the Froude-Reynolds phase space.

\begin{figure}
	\centerline{\includegraphics[scale=0.5]{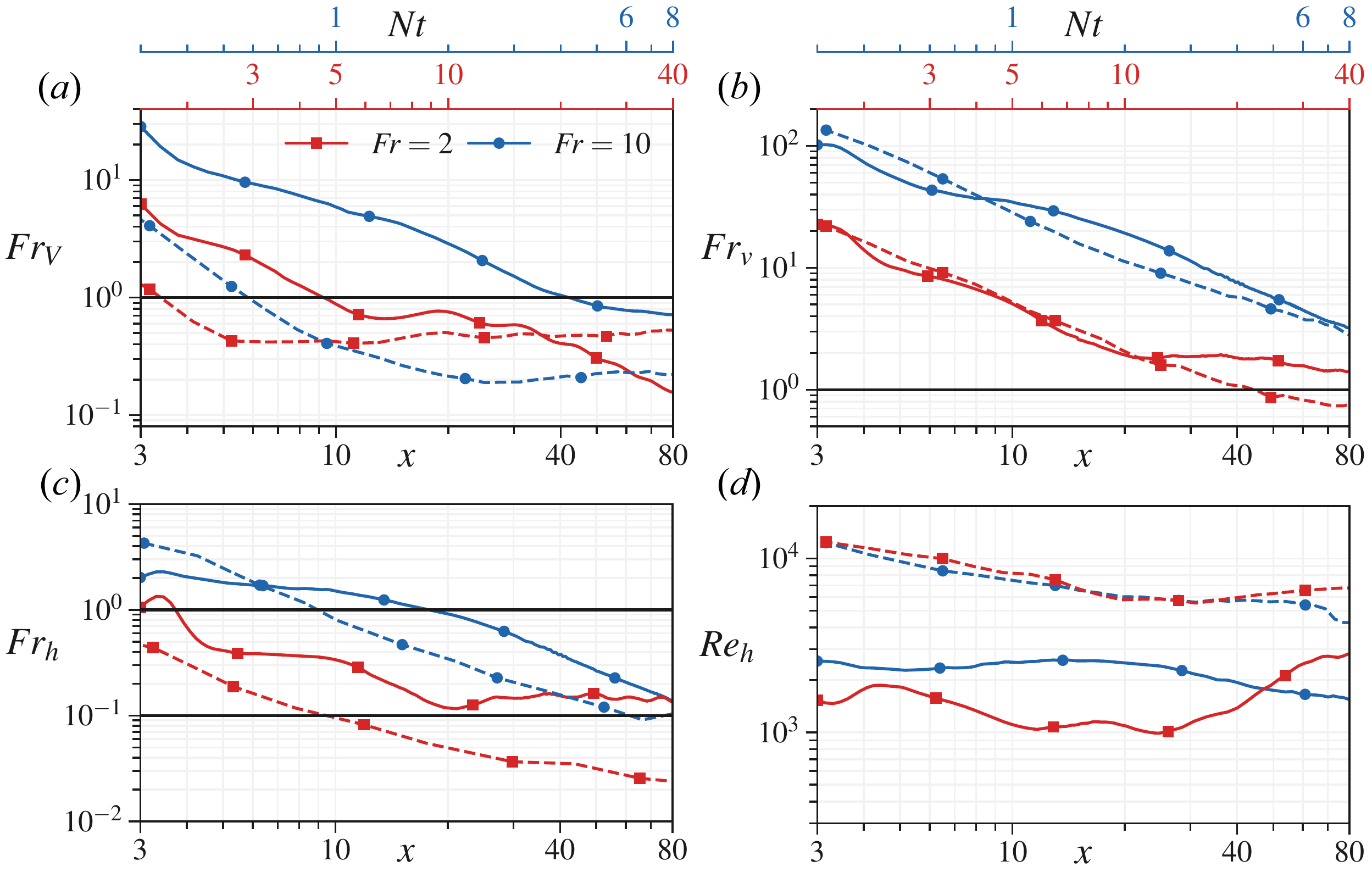}}
	\caption{Evolution of the spheroid (solid lines) and disk (dashed lines) wakes at $\Fro = 2$ (red) and $\Fro = 10$ (blue) in  the non-dimensional parameter space: (a) local vertical mean Froude number $\Fro_V$, (b) local vertical turbulent Froude number $\Fro_v$, (c) local horizontal turbulent Froude number $\Fro_h$, and 
	 (d) local horizontal Reynolds number.
	}
	\label{fig:NonDimNumbers}
\end{figure}

\begin{figure}
	\centerline{\includegraphics[scale=0.5]{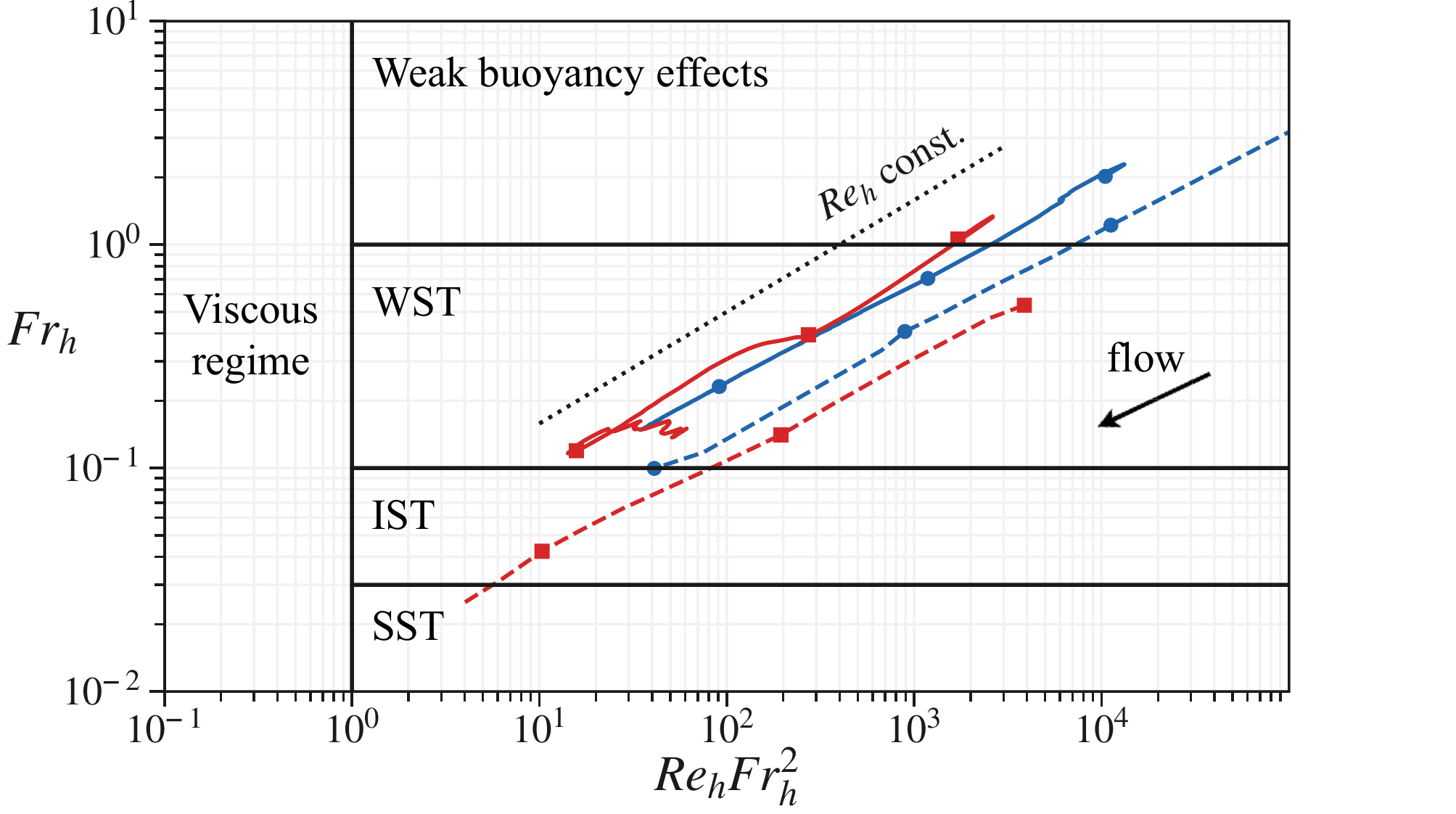}}
	\caption{Description of the trajectories of spheroid (solid lines) and disk (dashed lines) wakes  in $\Fro_h - \Rey_h\Fro_h^{2}$ phase space. Here, $\Fro = 2$ and $\Fro = 10$ are shown in red and blue, respectively. Dotted black line shows a $\Rey_h = \mathrm{const.}$ line in  phase space.}
	\label{fig:PhaseSpace}
\end{figure}


The mean vertical Froude number ($\Fro_V$) and the turbulent vertical and horizontal Froude numbers ($\Fro_v$, $\Fro_h$) are defined as 

\begin{equation}
\Fro_V=\frac{U_d}{2 N L_V},\qquad \Fro_v=\frac{u'_h}{N l_v},\qquad \Fro_h=\frac{u'_h}{N L_{Hk}},
\end{equation}
where $u_h'=(\langle u'^2_x\rangle+\langle u'^2_y\rangle)^{1/2}$ is the r.m.s. of the horizontal fluctuations and $l^2_v=\langle u'^2_x+u'^2_y\rangle/\langle(\partial_zu'_x)^2+ (\partial_zu'_y)^2)\rangle$ is a vertical turbulent length scale \citep{Riley2003}.

The mean Froude number is defined consistently with the global Froude number ($\Fro=U_\infty/ND$), this is, with the wake full height $2L_V$. The turbulent Froude numbers are defined with turbulence length scales in the energy-containing range. Since the horizontal turbulent integral length scale ($l_h$)  is not easy to compute in a spatially evolving flow, we use \revtwo{the TKE-based horizontal half wake width ($L_{Hk}$) as a surrogate following CS20. $L_{Hk}$ at a downstream location $x$ is calculated by $E^{T}_K(y=L_{Hk},z=0, x) = 0.5E^T_{K}(y=0,z=0, x)$.}

The mean vertical Froude number $\Fro_V$ (figure \ref{fig:NonDimNumbers}a) becomes $O(1)$ in both disk and spheroid wakes at the location at which the decay of $U_d$ slows down signaling the beginning of the NEQ regime, marked by $Nt \approx 1$ for the disk and $Nt \approx 3$ for the spheroid.  For the  $\Fro = 2$ cases, $\Fro_V$ in the spheroid wake starts dropping faster beyond $x \approx 40$ $-$ the streamwise location where the wake transitions from NEQ to Q2D regime.

The turbulent Froude numbers play an analogous role to those based on the mean. When their values become $O(1)$, turbulence starts being affected by buoyancy. $\Fro_v$ is defined using $l_v$ which has a significance to shear instability. Defined with $l_v$, $\Fro_v$ is proportional to $Ri^{-1/2}$, where $Ri$ is the Richardson number of the fluctuating shear ( \cite{Riley2003}, CS20).  When $\Fro_v$ becomes $O(1)$ (figure \ref{fig:NonDimNumbers}b) is also when the spanwise and vertical components of the TKE start showing anisotropy of the turbulence stress tensor and $E^T_{Ky}>E^T_{Kz}$ as shown in figure \ref{fig:energy_comps} for the $\Fro=2$ wakes. See also figure 8 of CS20.

\revtwo{The avid reader will notice that the four $\Fro_v=u'_h/Nl_v$ curves could collapse if plotted against $Nt$ instead of $x$. Indeed,  these curves collapse in the $(\Fro_v$, $Nt)$ plot for $1 < Nt < 10$ and the factor of $ 5$ present between the curves in figure \ref{fig:NonDimNumbers}(b) is simply the ratio of buoyancy frequency $N$ between $\Fro=2$ and $\Fro=10$. This result points to the fact that the spheroid and disk  wakes are at a sufficiently high $\Rey$ so that there is a range of small-scale wake turbulence with dynamics relatively unaffected by the large scales and, therefore, by neither the shape of the wake generator nor buoyancy.  $u'_h/l_v$ is a quantity that reflects such dynamics and thus, on dimensional grounds, evolves as $u'_h/l_v \sim t^{-1}$ so that $\Fro_v = u'_h/Nl_v \sim (Nt)^{-1}$. We note that  buoyancy does affect large scale wake features, which are affected by the body particularities -- hence the lack of collapse in $\Fro_V$. The values of the buoyancy Reynolds number $\Rey_b=\epsilon/\nu N^2$ and the Reynolds of the Taylor micro-scale $\Rey_\lambda=\sqrt{k} \lambda/\nu $ where $\lambda^2=15\nu u'^{2}_x/\epsilon$ (not included here for brevity) are very similar in magnitude for both disk and spheroid wakes at the two levels of stratification. This finding is consistent with the initial collapse of $\Fro_v$ when plotted against $Nt$.}


In figure \ref{fig:NonDimNumbers}(c), the $\Fro=10$ wakes of both disk and spheroid reach $\Fro_h \sim O(1)$ at $x\approx 10-20$, the location at which the $\{E^T_K\}$ starts deviating from its unstratified counterpart (figure \ref{fig:energy_comps}(a,b)). In the $\Fro=2$ spheroid wake, $\Fro_h<1$ throughout the domain and $\{E^T_K\}$ deviates from the unstratified decay from the very beginning in both disk and spheroid wakes. Overall, we find that $\Fro_h\sim O(1)$ is a good indicator of the deviation of $\{E^T_K\}$ from the unstratified counterpart. In contrast,  $\Fro_v \sim O(1)$ marks the location at which turbulence anisotropy between the vertical and the spanwise components starts growing. Figure  \ref{fig:NonDimNumbers}(d) shows the evolution of turbulence Reynolds number,  $\Rey_h = u'_hL_{Hk}/\nu$ with  $u'_h$ being the intensity of horizontal turbulent fluctuations. \revtwo{The value of $\Rey_h$ (figure \ref{fig:NonDimNumbers}d) changes slowly as the wake evolves, remaining at $O(10^4$) for the disk wake and at $O(10^3$) for the spheroid wake}.

A consolidated view of the evolution of the state of fluctuations is provided in phase space~\citep{Brethouwer2007, Zhou2019,deBKops2019,Chongsiripinyo2020}. In the phase-space portrait, one axis measures the buoyancy effect on the large scales through the turbulent Froude number ($\Fro_h$) and the other axis is a measure of  Reynolds number that is not $\Rey_h$ but one which accounts for buoyancy in addition to inertia. Ozmidov-scale eddies are the largest eddies unrestrained by buoyancy. The Ozmidov scales $l_o=(\varepsilon/N^3)^{1/2}$ and $u_o=(\varepsilon/N)^{1/2}$ lead to the definition of $\Rey_b=u_ol_o/\nu = \varepsilon/\nu N^2$ as the so-called buoyancy Reynolds number.  Another convenient measure of Reynolds number, which does not require explicit computation of the turbulent dissipation rate, is the buoyancy-weighted Reynolds number, $\Rey_h\Fro_h^2$~\citep{Billant2001,Riley2003}.  The Reynolds numbers based on buoyancy tend to decrease with downstream distance as buoyancy progressively increases in importance and limits the range of scales which are susceptible to 3D turbulent motions.  As long as $\Rey_h\Fro_h^2 > O(1)$, viscous effects do not dominate. 

Following CS20, figure \ref{fig:PhaseSpace} shows the evolution all four wakes in the $\Rey_h\Fro_h^2-\Fro_h$ phase space, where $\Rey_h = u'_hL_{Hk}/\nu$, $u'_h$ being the intensity of horizontal turbulent fluctuations. The flow evolves in the direction of the arrow from a state where buoyancy effects are weak (almost negligible)  to a region characterized by the presence of stratified turbulence. Within the state of stratified turbulence,  three different regimes can be demarcated: weakly, intermediately, and strongly stratified turbulence (WST, IST, and SST). Unlike the case of freely decaying turbulence, the mean velocity is also important here. Therefore, CS20 elected to distinguish between WST (where buoyancy begins to affect the mean velocity) and IST (where the turbulence anisotropy begins to be affected) as we do here. The SST regime is one where buoyancy effects on the large scales is very strong ($\Fro_h \ll 1$) but nevertheless the value of $\Rey_h\Fro_h^2$ is sufficiently large so that viscous effects are not dominant at the large scales.


The slope in the $\Rey_h\Fro_h^2 - \Fro_h$ plane is similar for both disk and spheroid wakes in the weak buoyancy and WST stages. \revtwo{The slope almost follows a line of constant $\Rey_h$, marked by a dashed line, signaling the slow evolution of $\Rey_h$ compared to that of $\Fro_h$}. However, there are significant differences in the way the spheroid wakes transverse the phase space. \revtwo{The spheroid wake starts out thinner than the disk wake, leading to larger local Froude numbers. Also, despite the body-based $\Rey$ being larger in the spheroid wake by a factor of two, the turbulence intensity in the disk wake is higher, hence the $\Rey_h$ difference observed in figure \ref{fig:NonDimNumbers}(d). This difference in $\Rey_h$ can be seen in the horizontal offset between the disk and spheroid lines in the phase-space representation.} 

Despite the larger spheroid body-based Reynolds number -- $\Rey = 10^5$ instead of $\Rey = 5\times 10^4$, turbulence in the spheroid wake is unable to access either the  the IST regime or the SST regime while the disk wake is able to access these regimes of stratified turbulence. Furthermore, the increase in TKE, which is not observed in the $\Fro = 2$ disk wake, reverses the trajectory of the $\Fro=2$ spheroid wake. To the authors' best knowledge, a reversing trend of phase-space trajectory has not been seen before in the stratified turbulence literature. These differences reveal that the phase space evolution, at least for $\Fro \leq 10$, depends on the features of the wake generator, e.g., its aspect ratio or the type of BL separation.

Based on the phase-space portrait alone, one may hastily conclude that stratified slender body wakes always experience a weaker buoyancy effect relative to bluff bodies. This is true for the $\Fro = 10$ case, at least for the limited $Nt$ simulation time, in terms of the relative amount of TPE (figure \ref{fig:energy_ratio}) and the vertical scale ($L_V$) deviation from the unstratified case -- note that both are smaller in the spheroid wake than in the disk. However, we have also shown that for the $\Fro = 2$ spheroid wake, compared to the disk wake, there is a much earlier onset of the Q2D regime. The mean and turbulence quantities are highly intertwined and the strong modulation of the mean flow by steady lee waves of the high aspect-ratio spheroid in the $\Fro =2$ case ultimately leads to an early entry into the Q2D regime. 
Hence, it is not entirely accurate to say that the slender body wakes are weakly affected by stratification. The present work calls for a need to generalize  the parameter space of $\{\Rey,\Fro\}$ in turbulent shear flows to account for the mean flow field (also possibly its instabilities) to build a more comprehensive understanding of buoyancy effects in shear flows. In the case of wakes, the shape of the body generator is brought into play through the mean flow as shown here.

\section{Discussion and final remarks}



The high-Reynolds number stratified wake of a slender body has been studied using a high-resolution hybrid simulation. The wake generator is a 6:1 prolate spheroid with a tripped turbulent boundary layer, the diameter-based Reynolds number is $\Rey=10^5$ and the Froude numbers, namely $\Fro=U_\infty/ND=\{2,10,\infty\}$, take moderate to  large values. By comparing the spheroid wake with the disk wake of \cite{Chongsiripinyo2020} (referred to as CS20), we are able to study  the influence of the wake generator - slender versus bluff - in the establishment and evolution of stratified wakes. 

The near wake of the 6:1 prolate spheroid with a turbulent boundary layer is characterized by
a small recirculation region ($\sim 0.1 D$). The recirculation region is surrounded by small-scale turbulence that emerges from the boundary layer and the flow does not show strong vortex shedding at the body\citep{Jimenez2010,Posa2016,Kumar2018,Ortiz2021}. As a result, the wake is much thinner and develops slower than the wake of a bluff body like the disk, which has a large recirculation region ($\sim 2 D$) and vortex shedding from the body \citep{Nidhan2020}. These body-dependent features of the near wake were recently shown to affect the decay of the far wake in environments {\em without }density stratification \citep{Ortiz2021}. In the present stratified simulations also we find substantial differences in the decay of the disk and spheroid wake. Particularly, we find that the starting locations of the non-equilibrium (NEQ) and the following quasi-2D (Q2D)  regions of wake deficit velocity depend on the wake generator.



At $\Fro=2 \approx (L/D)/\pi$, the wake of a 6:1 prolate spheroid is in a resonant state. The half wavelength of the lee waves is equal to the body length and, as a result, the flow separation and the wake are strongly modulated by the waves. Whereas previous works had described this regime in laminar-separation configurations of a sphere \citep{Hanazaki1988,Chomaz1992} and a 4:1 spheroid~\citep{Ortiz-tarin2019}, the present results show that the influence of the waves persists even at  high Reynolds numbers and with the separation of  a turbulent boundary layer. At $\Fro=2$, the  flow and the turbulence in the spheroid wake evolve very differently from the disk wake. Both the lack of strong shedding in the near wake \citep{Ortiz2021} and the strong modulation of the mean flow by the lee waves, lead to a wake with vertical  and horizontal profiles of mean velocity that depart strongly from Gaussian. These features are not observed in the disk wake at $\Fro =2$, which shows a vertically-squeezed Gaussian topology and a weak imprint of lee waves on the wake dimensions.

At $\Fro =2$, both disk and spheroid wakes transition to the NEQ regime at $Nt\approx \pi$. However the transition to the Q2D regime - with enhanced  wake decay relative to the NEQ regime - is very different; whereas the spheroid wake transitions at $Nt\approx 15$, the disk wake does not access the Q2D regime in a domain that spans $Nt\approx 60$. Other bluff bodies, e.g., the towed sphere~\citep{Spedding1997} show transition to the Q2D regime at $Nt \approx 50$, a location which is also delayed with respect to the spheroid wake. The early transition to the Q2D regime of the spheroid wake is driven by its strong modulation -- horizontal contraction and expansion of the wake width -- in response to the vertical contraction and expansion by the lee waves. This modulation has a particularly strong effect on the slender wake of a spheroid where the horizontal contraction is  
a large fraction of the wake width. The early start of the Q2D regime in the spheroid wake is accompanied by a sustained increase of turbulent kinetic energy (TKE), driven by an increase of the horizontal mean shear which acts on the turbulence of the separated boundary layer. The TKE increase is limited to the horizontal velocity with the spanwise  component being strongest, having  almost an order of magnitude larger energy than the vertical. Although coherent vortical structures and spanwise flapping are seen in the horizontal motion,  pancake eddies are incipient and not fully formed at  the end of the domain, $x/D = 80$.

At $\Fro=10$ also, there are differences between the disk and the spheroid wakes. Particularly in the spheroid wake, the beginning of the NEQ stage occurs later, at $Nt\approx 3$ instead of $Nt\approx 1$ ($x\approx 30$ instead of $x \approx 10$). The difference in the start of the NEQ can be attributed to the value of the local mean Froude number $\Fro_V=U_d/2NL_v$. As noted previously, the spheroid wake is thinner than the disk wake, the mixing in the near wake is weaker, and as a result the defect velocity in the intermediate wake is larger. These features increase the value of the spheroid wake local Froude number and delay the onset of the buoyancy effect that gives rise to the NEQ regime. Additionally, the analysis of the mean kinetic energy (MKE) transport terms shows that \revtwo{the  onset of buoyancy effect on the mean flow of both the disk and spheroid $\Fro=10$ wakes  is associated with 
the decreased energy transfer from MKE to TKE.} 

Taking the unstratified case as a base line,  the effect of buoyancy in the spheroid $\Fro=10$ wake is observed earlier (at $Nt \approx 1$) on the decay of the TKE than its effect (at $Nt \approx 3$) on the decay of $U_d$. In the spheroid wake at $\Fro = 10$, the transfer from TKE to TPE is responsible for the enhanced decay of TKE at  $Nt \approx 1$. \revtwo{On the other hand, the decrease in turbulent production at a farther downstream distance (compared to the disk $\Fro = 10$ wake) in the spheroid $\Fro=10$ wake 
is responsible for the slowed decay of the mean defect velocity at $Nt \approx 3$. The decrease in the production is caused by a reduction in the $\langle u'_xu'_z\rangle$ correlation \citep{JacobitzSV:1997,Spedding2002,Brucker2010}}. 
\revone{\cite{Meunier2004} compared the evolution far into the  stratified wake, up to $x \approx 8000$, among  several body shapes that also included a 6:1 prolate spheroid and a circular disk. The body Reynolds number was $\Rey = 5000$ \revone{and their diameter based Froude numbers were $\Fro = 4$ and $16$}. When normalized  using $D_{\mathrm{eff}}  = D \sqrt{C_{D}/2}$ instead of $D$, the evolution of the peak defect velocity of different wake generators exhibited approximate collapse for  $Nt \gtrapprox 50$ (see their figure 5b) with  a Q2D decay rate of $\sim$$x^{-0.75}$. Similarly, the wake width {in the horizontal} of different shapes approximately collapsed for $x/D_{\mathrm{eff}} > 400$ to exhibit a growth rate of $\sim$$x^{0.35}$. Since their 6:1 spheroid data starts from $x/D \approx 100$ (see their figure 5a), a direct comparison is not possible with our simulations that end at at $x/D = 80$.
	
In regard to the mean defect decay, a  major difference between \cite{Meunier2004} and our results is that the transition to Q2D power-law behavior appears earlier, around $Nt \approx 15$, in the $\Fro = 2$ spheroid wake relative to the the $\Fro = 2$ disk wake which does not transition to the Q2D decay rate until the end of the domain at $Nt = 62.5$. The value of $\Rey = 10^5$ in the spheroid wake is larger here and it is possible that the features that we have  linked to the early onset of the Q2D stage for the spheroid $\Fro = O(1)$ wake, i.e.,  the instability that leads to horizontal meanders and also the enhanced TKE production, are inhibited by viscous damping at the lower $\Rey$ of the experiments. 

The present simulations, both of the disk and the spheroid, do not extend into the very far wake regime reached by their experiments. Future  hybrid simulations or experimental work at higher $\Rey$ that probe the very far wake would clearly be useful. 
At any given $\Rey$ and $\Fro$, it will also be of interest to look at how tripping affects the wake evolution for the different body shapes. 
}



The simulations show that the buoyancy timescale $Nt$ alone is not sufficient to determine the state of the wake decay for both generators. However, we find that the value of the local turbulent and mean Froude numbers can be a good proxy to describe some aspects of the  wake state. For both disk and spheroid wakes, $\Fro_V=U_d/2NL_v$ becomes $O(1)$ at the location at which the decay of $U_d$ slows down; $\Fro_h=u'_h/N L_{Hk} \sim O(1)$ marks the location at which the area-integrated TKE of the stratified wake starts deviating from the unstratified case; and   $  \Fro_v=u'_h/N l_v \sim O(1)$  signals the location at which anisotropy between the different TKE components starts growing.

The buoyancy-weighted Reynolds number ($\Rey_h\Fro_h^{2}$) has been used widely in stratified flow as a convenient surrogate for the buoyancy Reynolds number ($\Rey_b = \varepsilon/\nu N^2$) since it displays similar trends during the flow evolution and the two quantities can be shown to be proportional using classical inviscid scaling of the turbulent dissipation rate. The surrogacy is true for the stratified wakes considered here except for the spheroid $\Fro = 2$ wake after its entry into the stage of Q2D wake decay. The horizontal fluctuation energy, therefore $\Fro_h$, increases owing to horizontal meanders  and flapping of the flow. However, $\varepsilon$ continues to decrease, albeit at a reduced rate relative to the NEQ regime. The value of $\Rey_b = O (10)$ is not high  in the Q2D regime realized here at $\Fro =2$. It remains to be seen if, in the Q2D regime at  even higher body-based Reynolds number, the equivalence between $\Rey_h{\Fro^2}_h$ and $\Rey_b$ is recovered and whether  the unusual upward trajectory seen here in $\{\Fro_h, \Rey_h\Fro_h^2\}$ phase space is also seen in $\{\Fro_h, \Rey_b\}$ space. The duration of the upward trajectory in phase space  until  the eventual  downward shift toward the viscous regime is also of interest.  



The differences between  bluff body (disk) and  slender body (6:1 spheroid)  wakes illustrate the difficulty of finding a universal scaling for the high-$\Rey$ stratified wake. The initial magnitude of $U_d$ for different wake generators and levels of stratification can be roughly scaled with the global $\Fro$ and the body drag coefficient \citep{Meunier2004}. However, the start and the duration of the NEQ regime  cannot be assumed to be independent of the wake generator. 
We find that rather than a particular $Nt$, the local  mean Froude number is a good proxy  for the onset of the NEQ regime in the mean defect velocity  and the values of  local turbulent Froude number provide guidance for the behavior of TKE, e.g., the onset of buoyancy effect  as well as the location at which the ratio of vertical to horizontal TKE starts decreasing. We are unable to  connect  Froude number  to the Q2D  regime transition of the wake.  More numerical and experimental work spanning different wake generators, different sources of turbulence including freestream turbulence,  and longer downstream distances will be instrumental in building a comprehensive picture of the effect of initial/boundary conditions on subsequent wake evolution.

\backsection[Acknowledgments]{Dr. K. Chongsiripinyo is thanked for useful discussions and for disk simulations. We thank the anonymous reviewers for their helpful comments.}
\backsection[Funding]{We gratefully acknowledge the support of Office of Naval Research Grant N00014-20-1-2253.}
\backsection[Declaration of interests] {The authors report no conflict of interest.}
\backsection[Author Contributions] {J.L.O.-T. and S.N. contributed equally to this paper and are co-first authors.}
\backsection[Author ORCIDs]{\\
J.L. Ortiz-Tarin https://orcid.org/0000-0001-8325-9381; \\
S. Nidhan        https://orcid.org/0000-0003-0433-6129; \\ 
S. Sarkar        https://orcid.org/0000-0002-9006-3173.}

\bibliographystyle{jfm}
\bibliography{references}
\end{document}